\documentstyle[ws-ijmpa,epsfig]{article}

\begin{document}

\thispagestyle{empty}
\def\catchline#1#2#3#4#5{\expandafter\def\expandafter\@clinebuf\expandafter
	{\@clinebuf\catchlinefont
	\noindent International Journal of Modern Physics A\par
	\noindent \copyright\ World Scientific Publishing Company\par
	}\relax\par
	}%

\markboth{P. Kanti}
{Black Holes in Theories with Large Extra Dimensions: a Review}

%
\catchline{}{}{}{}{}
%
\begin{flushleft}
{\small
International Journal of Modern Physics A\\
\copyright World Scientific Publishing Company}
\end{flushleft}

\bigskip

\title{BLACK HOLES IN THEORIES WITH\\ LARGE EXTRA DIMENSIONS: A REVIEW}

\author{\footnotesize PANAGIOTA KANTI}

\address{Theoretical Physics, University of Oxford, 1 Keble Road,\\
Oxford OX1 3NP, United Kingdom}

\maketitle \date{}

\smallskip


\begin{center}
{\begin{minipage}{4.6truein}
                \footnotesize
                 \par
We start by reviewing the existing literature on the creation of black holes
during high-energy particle collisions, both in the absence and in the presence
of extra, compact, spacelike dimensions. Then, we discuss in detail the
properties of the produced higher-dimensional black holes, namely the horizon
radius, temperature and life-time, as well as the physics that governs the
evaporation of these objects, through the emission of Hawking radiation.
We first study the emission of {\it visible} Hawking radiation on the
brane: we derive a {\it master} equation for the propagation of fields with
arbitrary spin in the induced-on-the-brane black hole background, and we
review all existing results in the literature for the emission of scalars,
fermions and gauge bosons during the {\it spin-down} and {\it Schwarzschild}
phases of the life of the black hole. Both analytical and numerical results
for the greybody factors and radiation spectra are reviewed and exact results
for the number and type of fields emitted on the brane as a function of the
dimensionality of spacetime are discussed. We finally study the emission of
Hawking radiation in the bulk: greybody factors and radiation spectra are
presented for the emission of scalar modes, and the ratio of the
{\it missing} energy over the {\it visible} one is calculated for different
values of the number of extra dimensions.
\end{minipage}}\end{center}

\noindent


\section{Introduction}	

A few years ago, in 1998, a novel idea was proposed according to which the
so-called {\it hierarchy problem} -- in other words, our difficulty in
answering the question of why the characteristic scale of gravity,
$M_P\sim 10^{19}$~GeV, is 16 orders of magnitude larger than the 
Electro-Weak scale,
$M_{EW} \sim 1$~TeV -- could be solved by assuming the existence of extra
dimensions in the Universe \cite{ADD,AADD}. The novelty in this idea was 
that the traditional picture of Planck-length-sized additional spacelike
dimensions ($\ell_P \simeq 10^{-33}$ cm) was abandoned, and the extra
dimensions could have a size as large as 1 mm. The upper bound on the size
of the proposed {\it Large Extra Dimensions} actually matched the smallest
length scale down to which the force of the gravitational interactions,
and thus their $1/r^2$ dependence, had been measured. If extra dimensions
of that size did exist, gravitational interactions would have a
completely different dependence on $r$ in scales smaller than 1 mm, 
however no gravitational experiment at that time could rule this out.

On the other hand, electromagnetic, weak and strong forces are indeed sensitive to
the existence of extra dimensions. If, for example, gauge bosons were allowed
to propagate in the extra-dimensional spacetime, their interactions would be
modified beyond any acceptable phenomenological limits unless the
size of the extra dimensions was smaller than $10^{-16}$\,cm. This problem
was resolved under the assumption that all particles experiencing this type
of interactions, in other words, all ordinary matter, is restricted to live
on a (3+1)-dimensional hypersurface, a {\it 3-brane}, that has a width along
the extra dimensions of, at most, the above order. The 3-brane, playing
the role of our four-dimensional world, is then embedded in the higher-dimensional
spacetime, usually called the {\it bulk}, in which only gravity can propagate. 

In this model, the large volume of the extra dimensions can help us solve, or
at least recast, the hierarchy problem: the traditional Planck scale, 
$M_{P}$, is only an effective energy scale derived from the fundamental
higher-dimensional one, $M_*$, through the following relation \cite{ADD}
\begin{equation}
M_P^2 \sim M_*^{2+n}\,R^n\,.
\end{equation}
In the above, it has been assumed that each one of the extra spacelike,
compact dimensions has the same size $R$. From the above, it becomes clear
that, if the volume of the compact space, $V \sim R^n$, is large, i.e if
$R \gg \ell_P$, then the  $(4+n)$-dimensional Planck mass, $M_*$, will be much
lower than the 4-dimensional one, $M_P$. If one chooses $M_*=M_{EW}$, then
the above expression provides a relation between the scale of gravity 
and the scale of particle interactions. 

The above idea makes use of geometrical features of the extra, compact
space in order to connect two completely different energy scales. The same
goal was achieved by an alternative, but similar idea, that was proposed a
year later \cite{RS} (for some early works on brane-world models and their
implications, see Refs. \refcite{Akama}-\refcite{Lykken}). According to this
alternative proposal,
the magnitude of the effective energy scale on the 3-brane, taken to be of
the order of the Electro-Weak scale, follows from the fundamental
higher-dimensional scale $M_P$ after being suppressed by an exponential
factor involving the distance of our observable brane from a hidden brane. 
In both models, however, the complete resolution of the hierarchy problem
would demand also an explanation of why the volume of the compact space, or
the inter-brane distance, has the value that leads to the observed ratio of
$M_P/M_{EW}$. 

In this review, we will concentrate on the scenario with Large Extra
Dimensions. A number of experiments and theoretical analyses have tried over
the years to put bounds on the size of extra, compact dimensions, and the
produced bounds are constantly updated. In the regime $r \ll R$, the extra
dimensions `open up' and Newton's law for the gravitational interactions
is modified assuming a $1/r^{n+2}$ dependence on the radial separation
between two massive particles.  Torsion-balance experiments which measure
the gravitational inverse-square law at short scales can provide limits
on the size of the extra dimensions, or equivalently on the value of the
fundamental scale $M_*$. On the other hand, since gravitons can propagate
both in the bulk and on the brane, massive Kaluza-Klein (KK) graviton states
can modify both the cross sections of Standard Model particle interactions
and astrophysical or cosmological processes. Assuming modifications that
are below the current observable limits puts bounds on the mass of the KK
gravitons, and consequently, on the size of extra dimensions. In Table 1,
we summarize some of those limits. The constraints from collider experiments,
although more accurate, are particularly mild, while the cosmological and
astrophysical ones are much more stringent; however, they contain large
systematic errors. If we ignore these errors, the latter type of constraints
exclude by far even the $n=3$ case, while low-gravity models with
$M_* \sim 1$~TeV are still allowed for $n \geq 4$.

\begin{table}[t]
\tbl{Current limits on the size of extra, compact dimensions}
{\begin{tabular}{@{}crr@{}} \toprule
Type of Experiment/Analysis &  $M_* \ge$ \hspace*{0.5cm} & $M_* \ge$
\hspace*{0.5cm}\\ \colrule
\begin{tabular}{l}Collider limits on the production \\
of real or virtual KK gravitons \cite{Delphi,Opal,CDF}\end{tabular} 
& 1.45 TeV ($n=2$) & 0.6 TeV ($n=6$)\\[4mm]  
Torsion-balance Experiments\cite{Hoyle}  & 3.5 TeV ($n=2$) &  \\[2mm]
Overclosure of the Universe \cite{Hall} & 8 TeV ($n=2$) & \\[2mm]
Supernovae cooling rate \cite{Cullen,Barger,Reddy,Reddy2} & 30 TeV ($n=2$)
&  2.5 TeV ($n=3$)\\[2mm]
Non-thermal production of KK modes \cite{Pospelov} & 35 TeV ($n=2$) & 
3 TeV ($n=6$)\\[2mm]
Diffuse gamma-ray background \cite{Hall,HR1,Hann} & 110 TeV ($n=2$) & 
5 TeV ($n=3$)\\[2mm]
Thermal production of KK modes~\cite{Hann} & 167 TeV ($n=2$) &  1.5 TeV ($n=5$)\\[2mm]
Neutron star core halo \cite{HR2} & 500 TeV ($n=2$) & 30 TeV ($n=3$) \\[2mm]
Neutron star surface temperature~\cite{HR2}& 1700 TeV ($n=2$) &  
60 TeV ($n=3$)\\[2mm]
BH absence in neutrino cosmic rays \cite{Feng} & & 1-1.4 TeV ($n \geq 5$)\\
\botrule
\end{tabular}}
\end{table}

If indeed present, the extra dimensions will inevitably change our notion
for the universe. The introduction of extra dimensions affects both
gravitational interactions and particle physics phenomenology, and leads to
modifications in standard cosmology. Already existing theories would need
to be extended or modified in order to accommodate the effects resulting
from the presence of extra dimensions (for an incomplete list of works on
the cosmological and phenomenological implications in theories with
large extra dimensions, see Refs. \refcite{DDG}-\refcite{Paul}). Similarly,
the properties and physics of black holes are also bound to change in the 
context of a higher-dimensional theory. 

As in the four-dimensional case, it seems natural to assume that, when matter
trapped on the brane undergoes gravitational collapse, a black hole is
formed which is centered on the brane and extends along the extra dimensions. 
If the horizon of the formed black hole is much larger than the size of
the extra dimensions, $r_H \gg R$, the produced black hole is effectively a
four-dimensional object. If, however, $r_H \ll R$, then this small black
hole is virtually a higher-dimensional object that is completely submerged
into the extra-dimensional spacetime. As we will see, these small black holes
have significantly modified properties compared to a four-dimensional black
hole with exactly the same mass $M_{BH}$: for example, they are larger,
colder and thus live longer compared to their four-dimensional analogues. 
Another striking consequence of the introduction of extra dimensions is that,
by lowering the Planck scale $M_*$ closer to the Electro-Weak scale, the
idea of the production of miniature black holes during high-energy scattering
processes, with trans-Planckian center-of-mass energy $\sqrt{s} \gg M_*$, now
becomes more realistic. Theoretical arguments have shown that the presence of
the extra dimensions facilitates further the production of black holes during
such collisions by increasing the production cross-section, thus leading to
striking consequences for the high-energy interactions of elementary particles
either at colliders or at cosmic rays.

The produced black holes are characterized by a 
non-vanishing temperature $T_H$, whose value is inversely proportional to
the horizon radius. They decay by the emission of Hawking radiation, i.e.
emission of elementary particles with rest mass smaller than $T_H$. This
is expected to be their most prominent observable signature with a 
characteristic thermal radiation spectrum and an almost blackbody profile. 
The non-trivial metric in the region exterior to the horizon of the black
hole creates an effective potential barrier which backscatters a part of
the outgoing radiation back into the black hole. The amount of radiation 
that finally reaches the observer at infinity depends on the {\it energy}
of the emitted particle, its {\it spin} and the {\it dimensionality} of
spacetime. The dependence on all the aforementioned parameters is encoded
into the expression of a filtering function, the `greybody factor'
$\sigma_{n}^{(s)}(\omega)$, which is present in the radiation spectrum.
The greybody factors can be important experimentally since they modify the
spectrum in the low- and intermediate-energy regime, where most particles
are produced, thus altering the characteristic spectrum by which we hope to
identify a `BH event'. In addition, as we will explain later in this review,
by studying the Hawking radiation emitted by this type of small black holes,
one would be able to `read' the total number of dimensions that exist in
nature.

A higher-dimensional black hole emits radiation both in the bulk and on
the brane.
According to the assumptions of the theory with Large Extra Dimensions, only
gravitons, and possibly scalar fields, can propagate in the bulk and thus,
these are the only types of fields allowed to be emitted in the bulk during
the Hawking evaporation phase. On the other hand, the emission on the brane
can take the form of scalar Higgs particles, fermions and gauge bosons. 
From the perspective of the brane observer, the radiation emitted in the bulk
will be a missing energy signal, while radiation on the brane may lead to 
experimental detection of Hawking radiation and thus of the production of
small black holes. Nevertheless, in order to have a clear picture of the
characteristics of the radiation spectrum on the brane, it is important to
know exactly how much energy is lost in the bulk. 

As we mentioned above, the greybody factor depends on the dimensionality
of spacetime; it also depends on whether the emitted particle is
brane-localized or free to propagate in the bulk. The greybody factor 
is actually the outgoing transmission cross-section associated with 
propagation in the aforementioned gravitational background, however,
due to the thermal character of the radiation spectrum, it is equal to
the incoming absorption cross-section \cite{Birrell}. Therefore, all we
need to do is to solve the equation of motion of a particle incident on
the background metric that describes the black hole, either
higher-dimensional or four-dimensional. After the absorption coefficient
is computed, the corresponding cross-section, and thus the greybody factor,
can easily follow. 

We need to stress here that the above semiclassical calculation of Hawking
emission is only reliable when the energy of the emitted particle is small
compared to the black hole mass, $\omega \ll M_{BH}$, since only in this
case is it correct to neglect the back reaction to the metric during the
emission process. This in turn requires that the Hawking temperature obeys
the relation $T_H \ll M_{BH}$, which is equivalent to demanding that the
black hole mass $M_{BH} \gg M_*$. As the decay proceeds and the mass of
the black hole decreases, inevitably this condition will break down during
the final stages of the evaporation process. Nevertheless, for black holes
of initial mass much larger than $M_*$ most of the evaporation process is
within the semi-classical regime. 

We will start this article by reviewing, in Section 2, the existing literature
on the creation of black holes during high-energy particle collisions, both
in four-dimensional and higher-dimensional spacetimes. In Section 3, we will
discuss the properties of the produced higher-dimensional black holes, namely
the horizon radius, the temperature and their life-time, and point out the
differences between them and their four-dimensional analogues. The physics
that governs the evaporation of these higher-dimensional objects, through
the emission of Hawking radiation, is covered in Section 4. Section 5 focuses
on the emission of Hawking radiation directly on the brane: we start with the
derivation of the {\it master} equation for the propagation of fields with
arbitrary spin in the induced-on-the-brane black hole background, and then
we present all existing results in the literature for the emission of scalars,
fermions and gauge bosons during the {\it spin-down} and {\it Schwarzschild}
phases of the life of the black hole; both analytical and numerical results
for the greybody factors and radiation spectra are presented as well as
exact results for the number and type of fields emitted on the brane as
a function of the dimensionality of spacetime. Section 6 deals with the
emission of Hawking radiation in the bulk and the question of the amount
of the missing energy: analytical and numerical results on the emission of
bulk scalar fields, including greybody factors and radiation spectra, are
reviewed, and the ratio of the missing energy over the `visible' one
emitted on the brane is presented for the Schwarzschild phase and for
different values of the number of extra dimensions. Our conclusions are
summarized in Section~7.


\section{High-energy Collisions and Black Hole Creation}

If extra dimensions do exist and the fundamental scale of gravity is much
lower than the traditional Planck scale $M_P$, then in the near future
we will witness collisions of particles with
trans-Planckian energies, i.e. energies larger than the fundamental
Planck scale $M_*$. Although, as we approach $M_*$, quantum gravity effects
become important for elementary particles, another semi-classical regime
opens up for center-of-mass energies of the colliding particles
$\sqrt{s} \gg M_*$: the products of such a collision would have a mass
much larger than the scale of quantum gravity and for these objects
(as for any macroscopic object) quantum gravity effects can be safely ignored. 

The high-energy scattering of particles and the nature of the products of
such collisions have been investigated in the framework of the General
Theory of Relativity \cite{khan,szekeres,dray,yurtsever,death},
String Theory \cite{gross,amati} and Quantum Gravity \cite{thooft,verlinde}.  
The usual approach followed is that the relativistic colliding particles 
(or black holes) may be described by two Aichelburg-Sexl gravitational
shock waves. In the limit of moving velocity equal to the speed of light,
the two particles can be considered massless and the curvature is zero
except on the null plane of their trajectory. If the impact parameter
$b$ is larger than the Schwarzschild radius $r_H$, that corresponds to the
center-of-mass energy of the two particles, elastic and inelastic processes
will in general take place between the two shock-waves accompanied by the
exchange of gravitons. If on the other hand, $b \leq r_H$, then, according to
the General Theory of Relativity and the ``hoop-conjecture" \cite{Thorne},
strong gravitational effects will dominate and a black hole will be formed. 
The production cross-section, in this high-energy limit,  is given by the
geometrical cross-section
\begin{equation}
\sigma \sim \pi b^2 \sim \pi r_H^2\,,
\end{equation}
i.e. by the ``target" area defined by the impact parameter.

As reported in Ref. \refcite{death}, Penrose found a lower bound on the mass
of the black hole produced during the collision of two particles moving
at the speed of light \cite{Penrose}. On the union of the two null planes,
that describe the trajectories of the two particles, an apparent horizon is
formed, with an area of $32 \pi \mu^2$ where $\mu$ is the energy of each
particle in the center-of-mass coordinate frame. This sets a lower bound on
the area of the event horizon (that according to the Cosmic Censorship
hypothesis~\cite{ellis} lies outside the apparent horizon) and thus on
the mass of the produced black hole; this is found to be
\begin{equation}
M_{BH} \geq {1 \over \sqrt{2}}\,2\mu\,,
\end{equation}
leading to the conclusion that at least 71\% of the initial center-of-mass 
energy
of the colliding particles will be bound into the black hole. The perturbative
analysis done subsequently in Ref. \refcite{death} determined the amount
of energy spent into gravitational radiation during the collision to be
16\% of the total energy, thus raising the estimate of the mass of the
black hole to 84\% of $\sqrt{s}$. The above results were derived under the
assumption of head-on collision ($b=0$) in four-dimensional spacetime.
An alternative study of the high-energy collision of a particle with a
Schwarzschild black hole in four dimensions \cite{Cardoso1} led to a similar
result: the emission of gravitational radiation is approximately 13\% of the
initial energy of the system; in the case of a rotating black hole, however,
this percentage can become as large as 35\%, but decreases as the impact
parameter increases.

After the theories with large extra dimensions and a low-scale gravity were
proposed \cite{ADD}, the idea of the creation of black holes from the
collision of particles with trans-Planckian energies was revived \cite{BF}.
The need for the update of the above lower limit on the mass of the black
hole for an arbitrary number of spacelike dimensions was obvious. Moreover,
realistic calculations of the rate of production of black holes during particle
collisions demanded the generalization of the same analysis for non-zero
impact parameters. By investigating the formation of a closed-trapped surface
around the colliding particles/shock-waves, the authors of Ref. \refcite{EG}
have shown that such surfaces do indeed form for both $D=4$ and $D > 4$ 
upon particular assumptions for the value of the impact parameter.
Of particular importance was their result that, in $D=4$, black holes form
if, and only if, $b<b_{max} \simeq 0.8 r_H$, a result which reduces
the value of the cross-section to $\sigma \simeq 0.65 \pi r_H^2$, and sets
the range of the black hole mass to 
\begin{equation}
M_{BH}=(0.71-0.45) \sqrt{s}\,, \qquad {\rm for} \quad b=\{0,b_{max}\}\,.
\end{equation}
Their higher-dimensional analysis, performed only for head-on collisions
($b=0$), revealed that the lower bound on the mass of the produced black hole
decreases also with the number of extra dimensions reaching $0.58 \sqrt{s}$
for $D=11$. In Ref. \refcite{Yoshino}, it was found that, for head-on
collisions and as $D$ increases, the circumference of the region into
which the mass $M$ must be compacted to produce a black hole is a decreasing
fraction of  $2 \pi r_H$. The same authors also showed that, for $b \neq 0$,
the value of $b_{max}$ that can produce a black hole increases with $D$ and
is given by: 
\begin{equation}
b_{max} \sim 2^{-1/(D-3)} r_H\,.
\end{equation}
This leads to the enhancement of the production cross-section, nevertheless,
the mass of the produced black hole will still be suppressed for large values
of $D$, in agreement with the results of Ref. \refcite{EG}. A recent
numerical analysis \cite{Berti} has produced some new estimates for the
energy lost in the form of gravitational radiation during a head-on collision
in a higher-dimensional spacetime: 13\% for $D=4$ to 8\% for $D=10$. While 
the former result is in agreement with Refs. \refcite{death} and
\refcite{Cardoso1}, the latter seems to disagree with the results produced
by the authors of Refs. \refcite{EG} and \refcite{Yoshino}. The apparent
disagreement can be resolved only under the assumption that a significant
part of the energy lost in the collision has a form different from that
of gravitational radiation.

The formation of
closed-trapped surfaces was studied analytically in Ref. \refcite{KV} for the
more realistic case of finite-front shock-waves, and was shown to indeed take
place for $D \geq 4$ and for an arbitrary impact parameter. The effect of the
angular momentum was looked at in Ref. \refcite{IOP}, where it was argued that
the production cross-section gets further enhanced when the spinning of the
black hole is taken into account. Finally, the evolution of the closed-trapped
surfaces in time was studied in Ref. \refcite{Vasilenko}.
 
But where and when these black holes may be produced? In the context of 
theories with large extra dimensions, the trans-Planckian energy regime may
lie slightly above the TeV scale. This raises the exciting possibility that
particle collisions with trans-Planckian energies may take place, in the near
future, at ground-based accelerators \cite{GT,dl,DE,giddings,GRW}, or they
may even already take place in the atmosphere of the earth
\cite{Goyal,FS,AG,EMR}. 
In the first case, accelerated particles (protons or nuclei) collide at
center-of-mass energies $\sqrt{s} > M_*$ at ground-based accelerators; if the
impact parameter is smaller than a critical value $b_c$, a higher-dimensional
black hole will be formed. The same holds in the case where 
highly-energetic cosmic ray particles with energies up to 100 TeV collide
with particles in the atmosphere of the earth. Neutrinos being scattered by
nuclei are expected to be the most effective source of black hole production
from cosmic ray particles due to the absence of any QCD-type contaminating
effects and small Standard-Model cross-sections.

The corresponding black hole production cross-sections are calculated in both
cases in the same way: first, any pair of partons $(i,j)$ that pass within
the Schwarzschild radius, i.e. have $b<b_c \simeq r_H(s)$, can lead to the
production of a black hole with cross-section $\sigma_{ij} \simeq \pi r_H^2(s)$,
where $\sqrt{s}$ is the center-of-mass energy of the colliding particles.
However, in a realistic collision, the colliding particles consist of more
than one partons, therefore, a summation must be made over all parton pairs
that carry enough energy to produce a black hole of a minimum mass $M_{min}$.
For this reason, the fraction of the center-of-mass energy that each parton 
carries must be taken into account via the use of the parton distribution
functions $f_i(x)$, that give the probability of finding a parton with a
fraction $x$ of the momentum of the colliding particle. The total
cross-section for the production of a black hole from the collision of two
particles $A$ and $B$ may then be written as \cite{GT,dl}
\begin{equation}
\sigma_{AB \rightarrow bh} (\tau_{m},s)= \sum_{i,j} \, \int_{\tau_m}^1 d\tau
\int_{\tau}^1 \frac{d x}{x}\, f_i (x)\,f_j(\frac{\tau}{x})\,\sigma_{ij}(\tau 
s)\,,
\end{equation}
where $\tau=x_i x_j$ is the parton-parton center-of-mass energy squared
fraction, and $\sqrt{\tau_m s}$ is the minimum center-of-mass energy necessary
for the creation of the minimum black hole mass, 
$M_{min} \simeq \sqrt{\tau_m s}$. In the
above, it has been assumed that both colliding particles are composite --
in case one of the particles is elementary (i.e. neutrinos) this expression
is simplified \cite{FS}. Early estimates found particularly large production
cross-sections of black holes: high-energy neutrinos scattered by nuclei at the
earth's atmosphere give a production cross-section, for a black hole with
$M_{min} =M_*=1$ TeV, which is two orders of magnitude larger than the
cross-section of any similar Standard Model process \cite{FS}. On the other
hand, proton-proton scattering at the LHC with $M_*=1$ TeV was found to lead
to a production cross-section of $10^5$ fb for $M_{min}=5$ TeV, a large
cross-section by new physics standards; for $M_{min}=10$ TeV, the cross-section
reduces to 10 fb. These estimates gave a substantial boost towards the further
study of the production and phenomenology of black holes in theories with
extra dimensions (see Refs. \refcite{Hossen}-\refcite{Barrau}). 

We should note here that the above production cross-sections decrease if one
takes into account the results of the studies on the formation of 
closed-trapped surfaces mentioned earlier in this section. By taking $b_c=r_H$ 
and not $b_c=b_{max}$, where $b_{max}$ is only a fraction of the Schwarzschild 
horizon, we obviously obtain an overestimate of the production cross-section. 
Moreover, assuming that the available energy for the production of black hole 
is the whole of the center-of-mass energy $\sqrt{s}$ is again an 
over-simplification since part of that energy is bound to be lost in the form 
of gravitational radiation during the collision \cite{death}. A claim for an 
additional suppression of this process was made \cite{voloshin} according to 
which an exponential suppression factor involving the Euclidean action of the 
system should be included. However further studies \cite{DE,KV,Solodukhin} 
argued that the black hole creation process from the collision of two particles 
was not classically forbidden and this suppression factor should not be taken 
into account. Subsequent analyses \cite{Rizzo,Hsu} calculated quantum 
corrections to the semi-classical cross-section and showed that these are 
indeed small. From a different perspective, the 
authors of Ref. \refcite{Das} have claimed that the use of the generalized
uncertainty principle leads to a radical increase in the minimum amount of
energy that is necessary for the creation of a black hole, thus rendering
unlikely the production of black holes at next-generation colliders such as
the Large Hadron Collider (LHC) with a center-of-mass energy of 14 TeV.
Recently, the validity of the description of the creation of a black hole
from the collision of a pair of Aichelburg-Sexl shock-waves was also
questioned \cite{Rychkov}, and the argument that strong-curvature and quantum
gravity effects may significantly alter the geometric cross-section estimate
was put forward.


\section{Properties of the Higher-Dimensional Mini Black Holes}

As we will see below, in the absence of extra dimensions, the creation 
of a semi-classical black hole demands extremely large center-of-mass energies,
which are far beyond our technical abilities. Reducing the mass of the black
hole down to accessible energy scales simply leads to an unnaturally small 
Schwarzschild radius, which can never be attained. On the other
hand, the presence of the extra dimensions facilitate the creation of 
black holes since it lowers the scale of quantum gravity, thus allowing
the production of semi-classical black holes at lower energies, and 
increases the corresponding Schwarzschild radius for a given
center-of-mass energy, thus making the black hole creation regime,
$b \leq r_H$, more easily accessible.  

Of particular importance, and simplicity, are the higher-dimensional black
holes that have a horizon radius much smaller than the size of the extra
dimensions, $r_H \ll R$. In this case, these mini black holes are completely
submerged into a $D$-dimensional spacetime that, to a very good approximation,
has one timelike and $D-1$ non-compact spacelike coordinates. If we further
assume that the produced black hole is spherically-symmetric, i.e.
non-rotating, the gravitational background around this black hole is
given by a generalized Schwarzschild line-element of the form \cite{Myers}
\begin{equation}
ds^2 = \left[1-\left(\frac{r_H}{r}\right)^{n+1}\right]\,dt^2 -
\left[1-\left(\frac{r_H}{r}\right)^{n+1}\right]^{-1}\,dr^2 - 
r^2 d\Omega_{2+n}^2\,,
\label{metric-n}
\end{equation}
where $n$ stands for the number of extra, spacelike dimensions that
exist in nature ($D=4+n$), and $d\Omega_{2+n}^2$ is the area of the 
($2+n$)-dimensional unit sphere given by
\begin{equation}
d\Omega_{2+n}^2=d\theta^2_{n+1} + \sin^2\theta_{n+1} \,\biggl(d\theta_n^2 +
\sin^2\theta_n\,\Bigl(\,... + \sin^2\theta_2\,(d\theta_1^2 + \sin^2 \theta_1
\,d\varphi^2)\,...\,\Bigr)\biggr)\,.
\label{unit}
\end{equation}
In the above, $0 <\varphi < 2 \pi$ and $0< \theta_i < \pi$, for 
$i=1, ..., n+1$. The line-element (\ref{metric-n}) can be easily shown to
satisfy the vacuum ($4+n$)-dimensional Einstein's equations. The black hole is
assumed to be bound on a 3-brane, our four-dimensional world, nevertheless,
the tension of the brane is assumed to be much smaller than the black hole
mass and thus it can be ignored in our analysis.

By using an analogous approach to the usual 4-dimensional Schwarzschild
calculation, i.e. by applying Gauss' law in the  $(4+n)$-dimensional spacetime,
we obtain the following relation between the horizon radius and the mass
$M_{BH}$ of the black hole \cite{Myers}
\begin{equation}
r_H= \frac{1}{\sqrt{\pi}M_*}\left(\frac{M_{BH}}{M_*}\right)^
{\frac{1}{n+1}}\left(\frac{8\Gamma\left(\frac{n+3}{2}\right)}{n+2}\right)
^{\frac{1}{n+1}}\,.
\label{horizon}
\end{equation}
We notice that, for $n \neq 0$, the relation between $r_H$ and $M_{BH}$ is
not linear anymore, and that it is the fundamental Planck scale $M_*$ that
appears in the expression of the horizon radius and not the four-dimensional
one $M_P$. The latter feature is the main reason for the fact that extra
dimensions facilitate the creation of low-mass black holes, as we will
shortly see. 

Before elaborating on this last point, we need to make another comment first:
in order to be able to ignore quantum corrections in our calculations and
study the produced black holes by using semi-classical methods, the mass of
the black hole must be, at least, a few times larger than the scale of
quantum gravity $M_*$. Therefore, if we assume that $M_* = 1$ TeV, a safe
limit for the mass of the produced black hole would be~\cite{GT,giddings}
$M_{BH} = 5$ TeV. By keeping fixed
the mass of the produced black hole, we may calculate the value of the
horizon radius as a function of $n$; these values are given in Table 2. 

\begin{table}[h]
\tbl{Black hole horizon radii for different values of $n$}
{\begin{tabular}{@{}c ccccccc@{}} \toprule
$n$ & 1 & 2 & 3 & 4 & 5 & 6 & 7 \\ \colrule
$r_H$ ($10^{-4}$ fm) & 4.06 & 2.63 & 2.22 & 2.07 & 2.00 & 1.99 & 1.99
\\ \botrule
\end{tabular}}
\end{table}

One may easily conclude, from the above, that during the collision of two
particles, with a center-of-mass energy $\sqrt{s} \geq 5$ TeV, a black hole 
may be formed if the particles pass within an area of radius $10^{-4}$ fm;
this is merely a sub-nuclear distance attainable at particle physics
experiments. On the other hand, in the absence of extra dimensions, the
lightest semi-classical black hole would have a mass of, at least, a few
times the four-dimensional Planck mass, $M_P \simeq 10^{16}$ TeV; the creation
of a black hole from a high-energy collision would then demand center-of-mass
energies higher than $M_P$, a requirement which is far beyond the reach of any
present and future accelerator. Overlooking for the moment the fact that a
four-dimensional black hole with $M_{BH} < M_P$ would not be a classical
object, we may ask what would be the value of the Schwarzschild radius for
an object with mass $M = 5$~TeV being produced in four dimensions in a
high-energy collision. We find that the Schwarzschild radius for such an
object has the extraordinary value of $r_H= 1.3 \times 10^{-50}$ m, i.e.
35 orders of magnitude smaller than the radius of the proton. 

The modified properties of a higher-dimensional, Schwarzschild-like black
hole, compared to those of a four-dimensional one with the same mass,
were first studied in Ref. \refcite{ADMR}. The fact that the Schwarzschild
radius in $D>4$ dimensions is larger than the one in $D=4$, for a given
mass $M_{BH}$, was first pointed out in there, and further implications
of the existence of extra dimensions on the temperature, life-time and
entropy of the black hole were investigated. We will now stop and look
in some detail at the temperature of a higher-dimensional black hole to
see where the difference from the four-dimensional case lies. The
temperature of a $(4+n)$-dimensional black hole is given by the
expression \cite{Myers}
\begin{equation}
T_H=\frac{(n+1)}{4\pi\,r_H}\,.
\label{temp}
\end{equation}
Let us assume again that $M_* = 1$ TeV, and that the produced black hole
has a mass $M_{BH} = 5$ TeV. By using Eq. (\ref{temp}) and the entries
of Table 2, we may easily calculate the temperature of the produced
black hole for different values of $n$; the results are displayed in 
Table 3.

\begin{table}[h]
\tbl{Black hole temperatures for different values of $n$}
{\begin{tabular}{@{}c ccccccc@{}} \toprule
$n$ & 1 & 2 & 3 & 4 & 5 & 6 & 7 \\ \colrule
\hspace*{1mm} $T_H$ (GeV) & 77 & 179 & 282 & 379 & 470 & 553 & 629
\hspace*{0.5mm}
\\ \botrule
\end{tabular}}
\end{table}

As we will shortly see, a black hole with a temperature $T_H$ emits
thermal radiation, the so-called Hawking radiation \cite{hawking}, through
the emission of ordinary particles. This leads to the decay of the black hole 
and finally to its evaporation. The radiation spectrum has a blackbody profile 
with the peak of the curve being at energies very close to its temperature. 
The temperature values displayed in Table 3 all lie in the GeV regime, which
is the energy range that present and next-generation experiments can probe. 
We may, thus, conclude that the presence of extra dimensions, not only 
facilitates the creation of a black hole at a high-energy collision, but also 
renders more likely the detection of their most prominent feature, the emitted 
Hawking radiation. 

We should note here that Hawking radiation emitted by larger, and thus
effectively four-dimensional, astrophysical black holes has never been
observed. Taking $M_{BH} \simeq 3 M_\odot$, which is the current upper
limit on the mass of neutron stars, and using Eq.~(\ref{temp}) with $n=0$,
we find that $T_H \simeq 10^{-12}\,{\rm eV} = 20$ nK; this is a very low
temperature corresponding to a very low energy frequency, or a very
large wavelength, that unfortunately cannot be detected. In the absence
of extra dimensions, the only black holes that could emit radiation
today at detectable frequencies, are primordial ones that were created at
the early universe and can have a much lower mass. For example, a black
hole with a mass $M_{BH}=10^{15}$ gr should give a Hawking radiation
spectrum with a peak in the area of 10-100 MeV; still, such a radiation
has not been yet observed. 

Combining Eqs. (\ref{horizon}) and (\ref{temp}), we may conclude that,
since a $(4+n$)-dimensional small black hole has a horizon radius much larger
than a four-dimensional one with the same mass, it will have a temperature
in $D>4$ which is much lower than the one in $D=4$ \cite{ADMR}. Indeed,
if a `quantum' black hole with $M_{BH}=5$ TeV had been allowed to exist in 
nature in $D=4$, its temperature would have been 30 orders of magnitude larger
than the entries in Table 3. The larger the temperature of the black hole
is, the faster its decay rate is -- through the emission of Hawking radiation
-- and thus the shorter its lifetime. Astrophysical black holes with masses 
$M_{BH} \geq 3 M_\odot$ emit radiation with an extremely small rate
\footnote{In addition, these black holes have a temperature much smaller
than the one of the Cosmic Microwave Background Radiation, $T=2.73\,{\rm K}=
2.3 \times 10^{-4}\,{\rm eV}$, therefore, they actually absorb radiation from
their environment instead of emitting.} and their lifetime, given by the
four-dimensional relation
\begin{equation}
\tau \sim \frac{1}{M_P}\,\biggl(\frac{M_{BH}}{M_P}\biggr)^3\,,
\label{life4}
\end{equation}
turns out to be much larger than the age of the universe. On the other hand,
the same formula tells us that tiny black holes (as the ones that might
have been created in the early universe in $D=4$) have an extremely short
life-time, due to their huge emission rate, and must have decayed
and evaporated long time ago. However, for small black holes, the presence
of extra dimensions modifies also their lifetime, and it is now
given by \cite{Myers}
\begin{equation}
\tau \sim \frac{1}{M_*}\,\biggl(\frac{M_{BH}}{M_*}\biggr)^{(n+3)/(n+1)}\,.
\label{life}
\end{equation}
The appearance of the low energy scale $M_*$ in Eq. (\ref{life}), instead of
$M_P$ as in Eq.~(\ref{life4}), leads to black hole lifetimes much longer
than the one in four dimensions. Nevertheless, for black holes with masses
in the area of a few TeV, the lifetime is still a tiny fraction of a second:
for $M_{BH} = 5$ TeV, the lifetime ranges from $1.7 \times 10^{-26}$ sec
(for $n=1$) to $0.5\times 10^{-26}$ sec (for $n=7$), while for $M_{BH} = 10$
TeV, the corresponding lifetime interval is from $1.6 \times 10^{-26}$ sec
(for $n=1$) to $1.2\times 10^{-26}$ sec (for $n=7$). Despite their short
lifetime, the fact that these small black holes might be created during
collisions on, or close to, the Earth's surface makes the possibility
of their observation a more realistic prospect.


\section{Evaporation of Higher-Dimensional Black Holes}

The emission of Hawking radiation is in fact compatible with the well-known
result of the General Theory of Relativity that nothing can escape from
inside the horizon of a black hole. The Hawking radiation can be conceived
as the creation of a virtual pair of particles just outside the horizon 
of the black hole: the particle with the positive energy
escapes to infinity while the antiparticle (the one with the negative
energy) falls into the BH, where it can exist as an ordinary particle. The
spectrum of the Hawking radiation coming from a black hole with temperature
$T_{H}$ is a thermal one with an {\it almost} blackbody profile. The flux
spectrum, i.e. the number of particles emitted per unit time, from a
higher-dimensional spherically-symmetric black hole of the type
(\ref{metric-n}), can be easily found by generalizing the corresponding
four-dimensional expression \cite{hawking} for a higher number of dimensions.
It is given by
\begin{equation}
\label{flux}
\frac{dN^{(s)}(\omega)}{dt} = \sum_{j} \sigma^{(s)}_{j,n}(\omega)\,
{1 \over \exp\left(\omega/T_{H}\right) \pm 1} 
\,\frac{d^{n+3}k}{(2\pi)^{n+3}}\,,
\end{equation}
where $s$ is the spin of the emitted degree of freedom and $j$ its angular
momentum quantum number. The spin statistics factor in the denominator is
$-1$ for bosons and $+1$ for fermions. For massless particles, $|k|=\omega$
and the phase-space integral reduces to an integral over the energy of the
emitted particle $\omega$. For massive particles, $|k|^2=\omega^2 -m^2$,
and the energy in the denominator now includes the rest mass of the
particle: this means that a black hole temperature $T_H \geq m$ is necessary
for the emission of a particle with mass $m$. We should note here that, as
the decay progresses, the black hole mass decreases and the Hawking
temperature rises. It is usually assumed that a quasi-stationary approach 
to the decay is valid -- that is, the black hole has time to come into 
equilibrium at each new temperature before the next particle is emitted.
We will make this assumption also here.

The power spectrum, i.e. the energy emitted per unit time by the black hole,
can be easily found by combining the number of particles emitted with the
amount of energy they carry. It is given by
\begin{equation}
\frac{dE^{(s)}(\omega)}{dt} = \sum_{j} \sigma^{(s)}_{j,n}(\omega)\,
{\omega  \over \exp\left(\omega/T_{H}\right) \pm 1}\,
\frac{d^{n+3}k}{(2\pi)^{n+3}}\,.
\label{power}
\end{equation}
Both expressions, Eqs. (\ref{flux}) and (\ref{power}), contain an additional
factor, $\sigma^{(s)}_{j,n} (\omega)$,  which does not usually exist in a
typical blackbody spectrum, or, more accurately, is just a constant standing
for the area of the emitting body. In contrast, this factor here depends on
the energy of the emitted particle, its spin and its angular momentum number.
It may therefore significantly modify the spectrum of the emitted radiation
and, for that reason, is called the `greybody' factor. Equally important is
the fact that this coefficient depends also on the number of extra dimensions
and therefore encodes valuable information for the structure of the spacetime
around the black hole including the dimensionality of spacetime.

The distortion of the blackbody spectrum, or in other words, the presence of
the greybody factor in the radiation spectrum of a black hole, can be
attributed to the following fact: any particle emitted by a black hole has
to traverse a strong gravitational background before reaching the observer at
infinity, unlike to what happens with a black body in flat spacetime. The
radiation spectrum is bound to depend on the energy of the propagating
particle and the shape of the gravitational barrier as these are the
parameters that will determine the number of particles that manage to reach
infinity. In order to illustrate the
above, let us assume, for example, that a scalar field of the form
$\phi\,(t,r,\theta_i,\varphi)=e^{-i\omega t}\,R_{\omega \ell}(r)\,
\tilde {Y}_\ell(\Omega)$, where $\tilde {Y}_\ell(\Omega)$ is the
$(4+n)$-dimensional generalization of the usual spherical harmonic
functions~\cite{Muller}, propagates in the background of Eq. (\ref{metric-n}).
Its radial equation of motion may then be written in a Schr\"odinger-like form
\begin{equation}
\left(-\frac{d^2 \,}{dy^2} + r^{2n+4}\left[-\omega^2
+\frac{\ell(\ell+n+1) h(r)}{r^2}\right] \right) R_{\omega \ell}(y) = 0\,,
\label{illus}
\end{equation}
in terms of the `tortoise' coordinate
\begin{equation}
y=\frac{\ln h(r)}{r_H^{n+1}\,(n+1)}\,, \qquad {\rm with} \quad
h(r) = 1-\biggl(\frac{r_H}{r}\biggr)^{n+1}\,.
\label{tortoise}
\end{equation}
The quantity inside square brackets in Eq. (\ref{illus}) gives the effective
potential barrier in the area outside the horizon of the black hole. It
clearly depends on all parameters mentioned above: $\omega$, $j$ (being
equal to the orbital angular momentum number $\ell$ in the case of a scalar
particle) and $n$. As a similar analysis will shortly show, it also depends
on the value of the spin $s$, for a non-zero spin particle. By simple
inspection, one may see that the barrier lowers for larger energy $\omega$,
while it rises for higher angular momenta $\ell$ -- the dependence, however,
on the number of extra dimensions is more subtle.

The greybody factor $\sigma^{(s)}_{j,n} (\omega)$ then stands for the
corresponding transmission cross-section for a particle propagating in 
the aforementioned background. This quantity can be determined by solving
the equation of motion of a given particle and computing the corresponding
absorption coefficient ${\cal A}^{(s)}_j$. Then, we may write~\cite{GTK} 
\begin{equation}
\sigma^{(s)}_{j,n} (\omega) = \frac{2^{n}\pi^{(n+1)/2}\,\Gamma[(n+1)/2]}
{ n!\,\omega^{n+2}}\,\frac{(2j+n+1)\,(j+n)!}{j !}\,|{\cal A}^{(s)}_j|^2\,.
\label{grey-n}
\end{equation}

It will be useful to rewrite the above expression for the greybody factor as
\begin{equation}
\sigma^{(s)}_{j,n} (\omega) = \frac{2^{n}}{\pi}\,
\Gamma\Bigl(\frac{n+3}{2}\Bigr)^2\,
\frac{A_H}{(\omega r_H)^{n+2}}\, N_j\,
|{\cal A}^{(s)}_j|^2\,,
\label{greyb}
\end{equation}
where $N_j$ is the multiplicity of states corresponding to the
same partial wave $j$, given for a $(4+n)$-dimensional spacetime
by \cite{Bander}
\begin{equation}
N_\ell= \frac{(2j+n+1)\,(j+n)!}{j! \,(n+1)!}\,,
\label{bulk-mult}
\end{equation}
and $A_H$ is the horizon area of the $(4+n)$-dimensional black hole
defined as
\begin{eqnarray}
A_H &=& 
r_H^{n+2}\,\int_0^{2 \pi} \,d \varphi \,\prod_{k=1}^{n+1}\,
\int_0^\pi\,\sin^k\theta_{k+1}\,
\,d\theta_{k+1} \nonumber \\[1mm]
&=& r_H^{n+2}\,(2\pi)\,\prod_{k=1}^{n+1}\,\sqrt{\pi}\,\,\frac{\Gamma[(k+1)/2]}
{\Gamma[(k+2)/2]}
\nonumber \\[1mm]&=& 
r_H^{n+2}\,(2\pi)\,\pi^{(n+1)/2}\,\Gamma\Bigl(\frac{n+3}{2}\Bigr)^{-1}\,.
\label{area-n}
\end{eqnarray}
From Eq. (\ref{greyb}), we may see that the greybody factor is indeed
proportional to the area of the emitting body, as in the case of blackbody
emission, nevertheless additional factors change, in principle,
this simple relation by adding an explicit dependence on $\omega$, $r_H$,
$j$ and $n$. We should also note that $\sigma^{(s)}_{j,n} (\omega)$ has the
same dimensions as $A_H$, therefore its dimensionality changes as the
number of extra dimensions $n$ varies. Normalizing its expression to
the area of the horizon of the $(4+n)$-dimensional black hole would
therefore be useful in comparing its values for different $n$.

Equation (\ref{power}), for energy emission in the higher-dimensional
spacetime, may take a simpler form when written in terms of the absorption
coefficient. Then, it reads
\begin{equation}
\frac{d E^{(s)}(\omega)}{dt} = 
\sum_{j} N_j\, |{\cal A}^{(s)}_j|^2\,{\omega  \over
\exp\left(\omega/T_{BH}\right) \pm 1}\,\,\frac{d \omega}{2\pi}\,.
\label{alter-bulk}
\end{equation}
In the above, we have assumed that the emitted particles are massless,
an assumption that will be made throughout this review. This simple,
alternative form of the power spectrum has led part of the community
to referring to the absorption probability $|{\cal A}^{(s)}_j|^2$ as the
greybody factor, this however will not be followed here.

As we mentioned above the greybody factor modifies the spectrum of emitted
particles from that of a perfect thermal blackbody \cite{hawking}.
Four-dimensional analyses for Schwarzschild \cite{page,unruh,sanchez} and
Kerr \cite{page2} black holes have determined, both analytically and
numerically, the greybody factors for particles of different spin. 
In the simplest case of a non-rotating, Schwarzschild black hole, geometric
arguments show that, in the limit of high energy $\omega$ (the geometrical
optics limit), $\Sigma_{j}\,\sigma^{(s)}_{j,n} (\omega)$ is a constant
independent of $\omega$ \cite{MTW} (for more details, see Sections 5 and 6). 
In that case, the spectrum is exactly like that of a blackbody for every
particle species independently of their spin $s$. The low-energy behaviour,
on the other hand, is strongly spin-dependent and energy-dependent, and 
the greybody factors are significantly different from the geometrical 
optics value \cite{page,sanchez,macG}.  The result is that both the power
and flux spectra peak at higher energies than those for a blackbody with the
same temperature. Finally, the spin dependence of the greybody factors
means that they are necessary to determine  the relative emissivities of
different particle types from a black hole. 

For a charged black hole, early works \cite{zaumen,gibbons,carter} showed
that the black hole will quickly discharge through a Schwinger-type
pair-production process. Moreover, the charge of the black hole could
affect the geometry of spacetime and thus the emission of uncharged
particles only for black holes with masses larger than $10^5\,M_\odot$.
A few years later, Page \cite{page3} showed that during the evaporation
phase, the charge of the black hole fluctuates. The electrostatic potential
between particle and antiparticle, together with the charge fluctuations,
reduces the emitted flux and power of charged particles, bringing it down 
to 7\% of the emission rate for similar but uncharged particles. In
the same work, the effect of the rest mass $m$ of the emitted particle
was studied and found that it also reduces the emission rate: for
electrons and muons, the reduction could be as large as 50\% for black
hole masses of ${\cal O}(10^{16}\,{\rm gr})$. 

In the light of the recent speculations of the possible creation of
higher-dimensional black holes during high-energy particle collisions,
the generalization of the existing four-dimensional analyses for the
calculation of greybody factors to a higher number of dimensions becomes
imperative. On the one hand, the existence of extra dimensions might
significantly change all that we know about the emission of particles
from a black hole, including the rate of emission and the type of particles
emitted. Moreover, if the dependence of the radiation spectrum
on the dimensionality of spacetime is strong enough, then, a possible
detection of such a spectrum can give us valuable information on the
dimensionality of spacetime.

If, eventually, small higher-dimensional black holes may be created in
high-energy particle collisions, then according to Refs. \refcite{GT} and
\refcite{giddings}, the produced black holes will go through a number of
phases before completely evaporating. These are:

\begin{itemlist}
\item
The {\it balding phase}\,: the black hole emits mainly gravitational radiation
and sheds the `hair' inherited from the original particles, and the asymmetry
due to the violent production process. 

\item
The {\it spin-down phase}\,: the typically non-zero impact parameter
of the colliding partons leads to black holes with some angular momentum
about an axis perpendicular to the plane. During this phase, the black
hole loses its angular momentum through the emission of Hawking radiation
and, possibly, through superradiance.

\item
The {\it Schwarzschild phase}\,: a spherically-symmetric black hole loses
energy due to the emission of Hawking radiation. This results in the gradual
decrease of its mass and the increase of its temperature.

\item
The {\it Planck phase}\,: the mass and/or the Hawking temperature approach
$M_*$ -- a theory of quantum gravity is necessary to study this phase in
detail.
\end{itemlist}

An important question related to the aforementioned life stages of the black
hole is how much energy is spent during each one of those phases. We have
already seen how the initial four-dimensional estimate of 16\% of the total
energy emitted in the form of gravitational radiation during the balding
phase, can go up to 55\% depending on the value of the impact parameter
and the dimensionality of spacetime. Crude estimates of the corresponding
percentages in four dimensions give 25\% and 60\% for the spin-down and
Schwarzschild phase \cite{GT,giddings}, respectively. However, it becomes
clear that exact analyses performed in the higher-dimensional spacetime,
in conjunction with the results derived already for the balding phase, can
radically change those numbers.  

In what follows, we will concentrate on
the second and third phase of the life of the black hole, that is the
spin-down and Schwarzschild phase. We will ignore the charge of the
black hole: although we expect the aforementioned 7\% figure, for the
emission rate of charged particles, to change with the addition of extra
dimensions, we anticipate that any increase would still allow us to
ignore the emission of charged particles, at least at first approximation.
We will also study the emission of only massless particles: the mass of the
black hole, being in the trans-Planckian regime of a few TeV's, leads 
to temperatures that are much larger than the rest masses of all
known particles (see Table 3). 

Another comment is in order here: the presence of the brane has so far
been ignored under the assumption that its tension is much smaller than
the mass of the black hole and thus it does not change the gravitational
background. Nevertheless, the brane plays another role, that of our
four-dimensional world where, according to the assumptions of the Theories
with Large Extra Dimensions, all ordinary particles (fermions, gauge bosons 
and Higgs fields) are localized. A $(4+n)$-dimensional black hole 
emits Hawking radiation both in the bulk and on the brane. Since only
gravitons, and possibly scalar fields, live in the bulk, these are the only
particles that can be emitted in the bulk.  On the other hand, the
emission of brane-localized modes include zero-mode scalars, fermions,
gauge bosons and zero-mode gravitons. From the observational point of view,
it is much more interesting to study the emission of brane-localized modes,
however, the emission in the bulk is equally important since it provides
answers to questions like how much energy is available for emission of
Standard Model particles on the brane. 

The brane-localized modes and bulk modes live in spacetimes of different
dimensionality. Whereas the bulk modes have access to the whole of the
$(4+n)$-dimensional spacetime, the brane modes live in a four-dimensional
slice of it, which is the projection of the higher-dimensional spacetime
on the brane. In order to study the emission of both types of modes, we
need to write down the corresponding equations of motion of those fields.
In the next section, we will first derive the {\it master equation} describing
the motion of a field with arbitrary spin $s$ in the background of a
higher-dimensional Kerr-like black hole projected onto our brane; then, we
will move on to review the so far derived results for the emission of Hawking 
radiation from rotating and non-rotating, uncharged $(4+n)$-dimensional
black holes on the brane. A similar review of the emission of Hawking
radiation in the bulk is left for Section 6.


\section{Master Equation and Hawking radiation on the Brane}

The gravitational background around a $(4+n)$-dimensional, rotating,
uncharged black hole was found by Myers \& Perry, and is given by the
following line-element~\cite{Myers}
\begin{eqnarray}
&~& \hspace*{-3cm}ds^2 = \biggl(1-\frac{\mu}{\Sigma\,r^{n-1}}\biggr) dt^2 +
\frac{2 a \mu \sin^2\theta}{\Sigma\,r^{n-1}}\,dt d\varphi 
-\frac{\Sigma}{\Delta}\,dr^2 -\Sigma\,d\theta^2 \nonumber \\[2mm]
\hspace*{2cm}
&-& \biggl(r^2+a^2+\frac{a^2 \mu \sin^2\theta}{\Sigma\,r^{n-1}}\biggr)
\sin^2\theta\,d\varphi^2 - r^2 \cos^2\theta\, d\Omega_{n},
\label{rot-metric}
\end{eqnarray}
where 
\begin{equation}
\Delta = r^2 + a^2 -\frac{\mu}{r^{n-1}}\,, \qquad
\Sigma=r^2 +a^2\,\cos^2\theta\,,
\label{Delta}
\end{equation}
and $d\Omega_n$ is the line-element on a unit $n$-sphere. The mass and
angular momentum (transverse to the $r \varphi$-plane) of the black hole
are given by
\begin{equation}
M=\frac{(n+2) A_{n+2}}{16 \pi G}\,\mu\,, \qquad
J=\frac{2}{n+2}\,M a\,,
\end{equation}
with $G$ being the $(4+n)$-dimensional Newton's constant, and $A_{n+2}$
the area of a $(n+2)$-dimensional unit sphere given by
\begin{equation}
A_{n+2}=\frac{2 \pi^{(n+3)/2}}{\Gamma[(n+3)/2]}\,.
\end{equation}

We will assume that the line-element (\ref{rot-metric}) describes successfully
a small, higher-dimensional black hole during its spin-down phase. At the
end of this phase, the angular momentum will be lost, and the same
line-element, in the limit $a \rightarrow 0$, will describe the following
spherically-symmetric Schwarzschild phase of the life of the black hole.
As mentioned earlier, in this section, we will be interested in the emission
of Hawking radiation on the brane, therefore we need to determine the
line-element in which the brane-localized modes propagate. This can be
found by fixing the values of the additional angular coordinates describing
the compact $n$ dimensions, and, then, the induced-on-the-brane line-element
is described again by Eq. (\ref{rot-metric}) but with the last term
omitted.

In order to derive a {\it master equation} describing the motion of a
field with arbitrary spin $s$ in the aforementioned projected background,
we need to make use of the Newman-Penrose formalism \cite{NP,Chandra}
that puts spinor calculus and gravitational quantities in the same 
framework. We first need to choose a tetrad basis of null vectors
$(\ell^\mu, n^\mu, m^\mu, \bar m^\mu)$, where $\ell$ and $n$ are real
vectors, and $m$ and $\bar m$ are a pair of complex conjugate vectors.
They satisfy the relations: ${\bf l} \cdot {\bf n}=1$, 
${\bf m}\cdot {\bf \bar m}=-1$, with all other products being zero.
Such a tetrad basis is given below:
\begin{eqnarray}
\ell^\mu=\Bigl(\frac{r^2+a^2}{\Delta},\,1, \,0,\,\frac{a}{\Delta}\Bigr)\,, &\quad &
n^\mu=\Bigl(\frac{r^2+a^2}{2\Sigma}, \,-\frac{\Delta}{2\Sigma}, \,0, 
\frac{a}{2\Sigma}\Bigr)\,, \nonumber \\[3mm]
m^\mu=\Bigl(ia \sin\theta, \,0, \,1, \,\frac{i}{\sin\theta}\Bigr)\,
\frac{1}{\sqrt{2} \bar\rho}\,, &\quad&
\bar m^\mu=\Bigl(-ia\sin\theta, \,0, \,1, \,\frac{-i}{\sin\theta}\Bigr)\,
\frac{1}{\sqrt{2} \bar\rho^*}\,,
\end{eqnarray}
where $\bar\rho=r+ia\cos\theta$. The $\lambda_{abc}$ coefficients, which are
used to construct the {\it spin coefficients}, are defined as:
\begin{equation}
\lambda_{abc}=(e_b)_{i,j}\Bigl[(e_a)^i (e_c)^j-(e_a)^j (e_c)^i \Bigr]\,,
\end{equation}
where $e_a$ stands for each one of the null vectors and $(i,j)$ denote the
components of each vector. To determine $\lambda_{abc}$ we will also need
the expressions of the null vectors with their index down, and these are:
\begin{eqnarray}
&&\ell_\mu=\Bigl(1,\,-\frac{\Sigma}{\Delta}, \,0,\,-a \sin^2\theta\Bigr)\,, 
\quad n_\mu=\Bigl(\frac{\Delta}{2\Sigma}, \,\frac{1}{2}, \,0, 
-\frac{a \sin^2\theta \Delta}{2\Sigma}\Bigr)\,, \nonumber \\[3mm]
&&m_\mu=\Bigl[ia \sin\theta, \,0, -\Sigma, \,-i\sin\theta (r^2+a^2)\Bigr]
\,\frac{1}{\sqrt{2} \bar\rho}\,, \quad
\bar m_\mu=(m_\mu)^*\,.
\end{eqnarray}
Then, the non-vanishing components of the $\lambda_{abc}$ coefficients are
found to be:
\begin{eqnarray}
&&\lambda_{122}=-\frac{1}{\Sigma^2}\,\Bigl(\frac{\Sigma}{2}\,
\partial_r \Delta - r\Delta\Bigr)\,, \quad 
\lambda_{132}=\frac{i \sqrt{2} a r \sin\theta}{\Sigma \bar \rho}\,, \quad
\lambda_{143}=\frac{1}{\bar\rho}\,,
\nonumber \\[1mm] && 
\lambda_{213}=-\frac{\sqrt{2} a^2 \cos\theta \sin\theta}{\Sigma \bar \rho}\,,
\quad \lambda_{243}=-\frac{\Delta}{2\Sigma \bar\rho}\,, \quad
\lambda_{314}=-\frac{2ia\cos\theta}{|\bar\rho|^2}\,, \nonumber \\[1mm] &&
\lambda_{324}=-\frac{i a \cos\theta \Delta}{\Sigma |\bar\rho|^2}\,,  \quad
\lambda_{334}=\frac{r\cos\theta + ia}{\sqrt{2} \bar\rho^2 \sin\theta}\,.
\end{eqnarray}
The above components must be supplemented by those that follow from the
symmetry $\lambda_{abc}=-\lambda_{cba}$ and the complex conjugates obtained
by replacing an index 3 by 4 (or vice versa) or interchanging 3 and 4 (when 
they are both present).

We may now compute the {\it spin coefficients} defined by
$\gamma_{abc}=(\lambda_{abc}+\lambda_{cab} -\lambda_{bca})/2$. Particular
components, or combinations, of the spin coefficients can be directly used
in the field equations \cite{NP,Chandra}. These are found to have the
values:
\begin{eqnarray}
\kappa=\sigma=\lambda=\nu=\epsilon=0\,, \quad \rho=-\frac{1}{\bar\rho^*}\,,
\quad \mu=-\frac{\Delta}{2\Sigma \bar\rho^*}\,,\quad 
\tau=-\frac{ia\sin\theta}{\sqrt{2}\Sigma}\,, \nonumber
\end{eqnarray}
\begin{equation}
\pi=\frac{ia\sin\theta}{\sqrt{2}\bar\rho^{*2}}\,, \quad 
\gamma=\frac{\partial_r\Delta}{4\Sigma}-\frac{\Delta}{2\Sigma \bar\rho^*}\,, 
\quad \beta=\frac{\cot\theta}{2\sqrt{2} \bar\rho}\,, \quad
\alpha=\pi-\beta^*\,. \label{spin}
\end{equation}
\smallskip

In what follows, we will also employ the Newman-Penrose operators:
\begin{eqnarray}
&& \hat D=\frac{(r^2+a^2)}{\Delta}\,\frac{\partial \,}{\partial t} +
\frac{\partial \,}{\partial r} + \frac{a}{\Delta}\,
\frac{\partial \,}{\partial \varphi}\,, \nonumber \\[2mm]
&&\hat \Delta = \frac{(r^2+a^2)}{2\Sigma}\,\frac{\partial \,}{\partial t} -
\frac{\Delta}{2\Sigma}\,\frac{\partial \,}{\partial r} +
\frac{a}{2\Sigma}\,\frac{\partial \,}{\partial \varphi}\,,\nonumber \\[2mm]
&&\hat \delta = \frac{1}{\sqrt{2} \bar\rho}\,\Bigl(ia\sin\theta\,
\frac{\partial \,}{\partial t}+\frac{\partial \,}{\partial \theta}+
\frac{i}{\sin\theta}\,\frac{\partial \,}{\partial \varphi}\Bigr)\,,
\label{oper}
\end{eqnarray}
and make use of the following field factorization:
\begin{equation}
\Psi_s(t,r,\theta,\varphi)= e^{-i\omega t}\,e^{i m \varphi}\,R_{s}(r)
\,S^{m}_{s,j}(\theta)\,,
\label{facto}
\end{equation}
where $S^{m}_{s,j}(\theta)$ are the so-called spin-weighted spheroidal
harmonics. For $s=0$, these angular eigenfunctions reduce to the spheroidal
harmonics \cite{Flammer}, while for $a \omega=0$, they take the form of the
spin-weighted spherical harmonics \cite{goldberg}. In what follows, we will
address separately the equation of motion of each type of field (gauge
bosons, fermions and scalars) in the background of the projected, Kerr-like
black hole.

\bigskip \noindent
{\bf Gauge Bosons ($s=1$)}.
In the Newman-Penrose formalism, there are only three `degrees of freedom'
for a gauge field, namely $\Phi_0=F_{13}$, $\Phi_1=(F_{12}+ F_{43})/2$ and
$\Phi_2=F_{42}$, in terms of which the different components of the Yang-Mills 
equation for a massless gauge field are written as:
\begin{eqnarray}
(\hat D-2 \rho)\,\Phi_1 - (\hat \delta^* + \pi -2 \alpha) \,\Phi_0 &=& 0\,, 
\label{B1}\\[1mm]
(\hat \delta -2\tau)\,\Phi_1 - (\hat \Delta + \mu-2 \gamma) \,\Phi_0 &=& 0\,, 
\label{B2}\\[1mm]
(\hat D-\rho)\,\Phi_2 - (\hat \delta^* + 2\pi)\,\Phi_1 &=& 0\,, \label{B3}\\[1mm]
(\hat \delta-\tau +2 \beta)\,\Phi_2 - (\hat \Delta +2 \mu) \,\Phi_1 &=& 0\,, 
\label{B4}
\end{eqnarray}
where $\hat \delta^*$ stands for the complex conjugate of $\hat \delta$.
One of the main difficulties in writing down equations of motion for fields
with a non-zero spin, and thus with more than one components, is to decouple
the differential equation for a particular component from the remaining
ones. This is not always possible but for the axially-symmetric Kerr-like
black hole that we consider here it is. The situation actually is very
similar to the pure four-dimensional case studied by Teukolsky 
\cite{Teukolsky}, where it was shown that at least the radiative components,
that carry all the information for the propagating field, do indeed decouple
\cite{TP}. Rearranging Eqs. (\ref{B1})-(\ref{B2}) and using the identity
\begin{equation}
(\hat D - 2 \rho -\rho^*)(\hat \delta -2 \tau)\,\Phi_1=
(\hat \delta -2 \tau) (\hat D-2 \rho)\,\Phi_1\,,
\end{equation}
we obtain the following decoupled equation for $\Phi_0$
\begin{equation}
(\hat D - 2 \rho -\rho^*)(\hat \Delta + \mu -2 \gamma)\,\Phi_0-
(\hat \delta -2 \tau) (\hat \delta^* +\pi -2 \alpha)\,\Phi_0=0\,.
\end{equation}
Using the explicit forms of spin coefficients and operators, Eqs.
(\ref{spin}) and (\ref{oper}), as well as the factorized ansatz
(\ref{facto}), the above can be conveniently separated into an angular
equation,
\begin{eqnarray}
&& \hspace*{-1.3cm} \frac{1}{\sin\theta}\,\frac{d \,}{d \theta}\,
\biggl(\sin\theta\,\frac{d S^m_{1,j}}{d \theta}\biggr)
+ \biggl[ -\frac{2 m \cot\theta}{\sin\theta} -\frac{m^2}{\sin^2\theta}
+ a^2 \omega^2 \cos^2\theta \nonumber \\[1mm]
&& \hspace*{3.2cm} - 2 a \omega \cos\theta + 1 - \cot^2\theta
+ \lambda_{1j} \biggl] S^m_{1,j}(\theta)=0\,,
\end{eqnarray}
where $\lambda_{1j}$ is a separation constant, and a radial equation,
\begin{equation}
\frac{1}{\Delta}\,\frac{d \,}{d r}\biggl(
\Delta^2\,\frac{d R_1}{d r}\biggr) + \biggl[
\frac{K^2-iK \partial_r\Delta}{\Delta} + 4 i \omega r 
+ (\Delta'' -2) -\Lambda_{1j}\biggr]\,R_1(r)=0\,,
\end{equation}
where we have defined:
\begin{equation}
K=(r^2+a^2)\,\omega -a m\,, \qquad \Lambda_{sj}=\lambda_{sj} + a^2 \omega^2
-2 a m \omega\,. \label{lambda}
\end{equation}

\bigskip \noindent
{\bf Fermion Fields ($s=1/2$)}. For a massless two-component spinor field,
the Dirac equation can be written as:
\begin{eqnarray}
(\hat \delta^*-\alpha +\pi)\,\chi_0 &=& (\hat D-\rho)\,\chi_1\,, \\[1mm]
(\hat \Delta + \mu -\gamma)\,\chi_0 &=& (\hat \delta + \beta-\tau)\,\chi_1\,.
\end{eqnarray}
Performing a similar rearrangement as in the case of bosons and using
the identity
\begin{equation}
(\hat \delta-\alpha^* -\tau+\pi^*)(\hat D-\rho)\,\chi_1 = 
(\hat D-\rho-\rho^*)(\hat \delta + \beta-\tau)\,\chi_1\,,
\end{equation}
we find that $\chi_1$ is decoupled leaving behind an equation for $\chi_0$
\begin{equation}
(\hat D-\rho-\rho^*)(\hat \Delta + \mu -\gamma)\,\chi_0-
(\hat \delta-\alpha^* -\tau+\pi^*)(\hat \delta^*-\alpha +\pi)\,\chi_0=0\,. 
\end{equation}
The latter can be explicitly written as a set of angular and radial equations,
having the form
\begin{eqnarray}
&& \hspace*{-1cm} \frac{1}{\sin\theta}\,\frac{d \,}{d \theta}\,
\biggl(\sin\theta\,\frac{d S^m_{1/2,j}}{d \theta}\biggr)
+ \biggl[ -\frac{m \cot\theta}{\sin\theta} -\frac{m^2}{\sin^2\theta}
+ a^2 \omega^2 \cos^2\theta \nonumber \\[2mm]
&& \hspace*{3.2cm}  - a \omega \cos\theta + \frac{1}{2} - 
\frac{1}{4}\,\cot^2\theta + \lambda_{\frac{1}{2}j} \biggl] 
S^m_{1/2,j}(\theta)=0\,,
\end{eqnarray}
and 
\begin{equation}
\frac{1}{\sqrt{\Delta}} \frac{d \,}{d r}
\biggl(\Delta^{3/2} \frac{d R_{1/2}}{d r}\biggr) + \biggl[
\frac{K^2-iK (\partial_r\Delta)/2}{\Delta} + 2 i \omega r 
\,+ \frac{1}{2}(\Delta'' -2) -\Lambda_{\frac{1}{2}j}\biggr] R_{1/2}(r)=0,
\end{equation}
respectively, with the same definitions for $K$ and $\Lambda_{sj}$
as before.

\bigskip \noindent
{\bf Scalar Fields ($s=0$)}. In the case of a scalar field, its equation
of motion can be determined quite easily by evaluating the double covariant
derivative $g^{\mu\nu} D_\mu D_\nu$ acting on the field. As in the purely
four-dimensional case \cite{Brill}, the use of the factorized ansatz
(\ref{facto}) leads, similarly to the previous cases, to the following pair
of separated equations
\begin{equation}
\frac{1}{\sin\theta}\,\frac{d \,}{d \theta}\,\biggl(\sin\theta\,
\frac{d S^m_{0,\ell}}{d \theta}\,\biggr) + \biggl[-\frac{m^2}{\sin^2\theta}
+a^2 \omega^2 \cos^2\theta + \lambda_{0\ell} \biggr]\,S^m_{0,\ell}=0\,,
\end{equation}
\smallskip
\begin{equation}
\frac{d \,}{dr}\,\biggl(\Delta\,\frac{d R_0}{dr}\biggr) +
\Bigl(\frac{K^2}{\Delta} - \Lambda_{0\ell}\Bigr) R_0(r) =0\,,
\label{scalar}
\end{equation}
where $Y^m_{0,\ell}=e^{i m \varphi}\,S^{m}_{0,\ell}(\theta)$ are now the
spheroidal wave functions \cite{Flammer}.

\bigskip \smallskip 
Combining all the above equations derived for bosons, fermions and scalar
fields, we may now rewrite them in the form of a {\it master equation} valid
for all types of fields. The radial equation then takes the form:
\begin{equation}
\Delta^{-s} \frac{d \,}{dr}\,\biggl(\Delta^{s+1}\,\frac{d R_s}{dr}\,\biggr) +
\biggl(\frac{K^2-is K \partial_r\Delta}{\Delta} + 4 i s \omega r 
+ s\,(\Delta'' -2) -\Lambda_{sj}\biggr) R_s(r)=0\,,
\label{master1}
\end{equation}
while the angular equation reads
\begin{eqnarray}
&& \hspace*{-1cm} \frac{1}{\sin\theta}\,\frac{d \,}{d \theta}\,
\biggl(\sin\theta\,\frac{d S^m_{s,j}}{d \theta}\,\biggr) + 
\biggl[-\frac{2 m s \cot\theta}{\sin\theta} - \frac{m^2}{\sin^2\theta}
+ a^2 \omega^2 \cos^2\theta \nonumber \\[2mm]
&& \hspace*{3.2cm}  - 2 a s \omega \cos\theta  + s - s^2 \cot^2\theta 
+ \lambda_{sj} \biggr]\,S^m_{s,j}=0\,.
\label{master2}
\end{eqnarray}
The latter equation is identical to the one derived by Teukolsky
\cite{Teukolsky} in the case of a rotating, Kerr black hole. The radial
equation differs by the extra factor $s\,(\Delta''-2)$ due to the fact that
for our metric tensor this combination is not zero, contrary to what happens
in the case of the 4-dimensional Kerr, or Schwarzschild, metric. 

In order to solve the radial equation (\ref{master1}), we need to know
the value of the constant term $\Lambda_{sj}$ defined in Eq. (\ref{lambda}),
or alternatively the value of the separation constant $\lambda_{sj}$ that
appears in the angular equation (\ref{master2}). Actually, the separability
of radial and angular parts in the equation of motion of a field in a
Kerr-like black hole background comes with a price:
the separation constant is a complicated function of both the energy
$\omega$ of the particle and the angular momentum parameter $a$ of the
black hole. An analytic form can be found, for any value of $s$, in the
limit of $a \omega \rightarrow 0$, in which case we may write
\cite{Staro,Fackerell,Seidel}
\begin{equation}
\lambda_{sj}= -s(s+1) + \sum_k f_k^{jms} (a \omega)^k=
j(j+1)-s(s+1) -\frac{2 m s^2}{j(j+1)}\,a \omega + ...\,,
\end{equation}
where terms higher than linear have been suppressed due to their complexity.
It becomes therefore clear that any attempt to solve analytically the
radial equation (\ref{master1}) must necessarily take place in the
low-energy and low-momentum limit. 

Finally, we may simplify further the radial equation (\ref{master1}) by
making the redefinition $R_s=\Delta^{-s} P_s$. In that case, the
$\Delta''$-term disappears and we obtain:
\begin{equation}
\Delta^{s}\,\frac{d \,}{dr}\,\biggl(\Delta^{1-s}\,\frac{d P_s}{dr}\,\biggr) +
\biggl(\frac{K^2-is K \partial_r\Delta}{\Delta} + 4 i s \omega r 
- \tilde \Lambda_{sj} \biggr)\,P_s(r)=0\,,
\label{master3}
\end{equation}
where now  $\tilde \Lambda_{sj}=\Lambda_{sj} +2 s$. 

Having derived the equation of motion for fields with arbitrary spin $s$
propagating in the four-dimensional spacetime induced on the 3-brane, we
can now move on to discuss the emission of Hawking radiation directly on
the brane during the spin-down and Schwarzschild phase of a small,
higher-dimensional black hole.

\subsection{Emission of Hawking radiation during the spin-down phase}

The study of the emission of Hawking radiation on the brane during the
spin-down phase has not been as thorough as one would hope. The only
such attempt has been made by the authors of Ref. \refcite{IOP}, who
independently performed the derivation of the {\it master equation}
described earlier \footnote{A typographical error appears in
the published version of Ref. \refcite{IOP}: the coefficient $s$ that
should have multiplied the $\Delta''$-term [or $n(n-1)\mu r^{n-1}$,
in their notation] is constantly missing from their equations for
scalars (A5), fermions (A23) and gauge bosons (A26), as well as from
the master equation (A28). Nevertheless, the results presented in
Ref. \refcite{IOP} are still robust since they were derived
for the particular case of $n=1$ for which this term vanishes trivially.}. 
The master equation was solved analytically in the limit of small
$a \omega$ and for the particular case of only one extra compact dimension,
i.e. $n=1$. A well-known approximation method
\cite{Staro,page,unruh,sanchez,Mathur}$^-$\cite{Lee} was used according
to which the asymptotic solutions at infinity and close to the horizon are
found, and then matched in an intermediate zone in order to construct the
complete analytic solution. The same method had been used earlier in
Ref. \refcite{kmr1} to compute the spectrum of Hawking radiation of scalar
fields emitted both on the brane and in the bulk, and later, in Ref.
\refcite{kmr2}, for the study of the emission of fermions and gauge bosons
on the brane; both analyses were done for the Schwarzschild phase of
the black hole, therefore, a description of the aforementioned approximation
method will be given in the next subsection. 

In Ref. \refcite{IOP}, the greybody factors, as well as the radiation
spectra, for the emission of scalars, fermions and gauge bosons on the
brane during the spin-down phase were derived for $n=1$. For various
values of the angular momentum parameter $a$, the radiation spectra
were plotted as a function of the energy $\omega$ of the emitted particle.
The presented results reveal a suppression of the power spectrum for 
scalar fields, as the black hole angular momentum increases, both at low
and high energies. On the other hand, in the radiation spectra for fermions
and gauge bosons, we observe an enhancement at low energies that turns to
a suppression at high energies. In all cases, the power emission curves,
for all values of $a$ and $s$, lie far below the ones produced by using
the high-energy geometrical optics limit value of the greybody factors.

However, these analytic results were derived in the limit of small
$a \omega$, therefore, they are bound to break down at high energies.
The high energy behaviour of the power spectra presented are thus not
trustable and an exact (numerical) analysis is necessary to determine
the complete spectrum. By comparing the results, for $a=0$ and $n=1$,
produced in Ref. \refcite{IOP} with the ones derived in the four-dimensional
case \cite{page}, we conclude that the power spectrum of all
fields seems to be enhanced. However, for $a>0$, the suppression caused
by the black hole angular momentum complicates the picture and the
comparison with the four-dimensional case is not clear. Similarly, no
conclusions can be drawn on the behaviour of the radiation spectra
as $n$ increases. Apart from an exact analysis, valid at
all energy regimes, a complete study of the dependence of the emission
spectra on the number of extra dimensions, is clearly necessary. Both
of these tasks have been successfully performed in the case of the
Schwarzschild phase~\cite{HK}, and the results are reviewed in the next
subsection.


\subsection{\hspace*{-2.5mm}
Emission of Hawking radiation during the Schwarzschild phase}

In the four-dimensional case, the most optimistic estimates for the amount
of the initial energy of the colliding particles lost in the balding
($\sim$16\%) and
spin-down ($\sim$25\%) phases leave behind approximately 60\% of the total
energy to be spent during emission of Hawking radiation in the Schwarzschild
phase. Although we now know that these numbers change with the addition of
extra dimensions, we believe that the Schwarzschild phase will still be the
longer one and will account for the greatest proportion of the mass loss in
the life of a small, higher-dimensional black hole. Contrary to the case
of the spin-down phase, the emission of Hawking radiation on the brane
during the Schwarzschild phase has been thoroughly studied, both
analytically \cite{kmr1,kmr2} and numerically \cite{HK}. In what follows
we present a review of both types of results (see also \cite{kanti2}).

A higher-dimensional black hole with vanishing angular momentum induces
on the 3-brane the following spherically-symmetric, Schwarzschild-like 
line-element
\begin{equation}
ds^2=- h(r)\,dt^2 + h(r)^{-1}\,dr^2 + r^2\,(d\theta^2 + 
\sin^2\theta\,d\varphi^2)\,,
\label{metric-4}
\end{equation}
where still
\begin{equation}
h(r) = 1-\biggl(\frac{r_H}{r}\biggr)^{n+1}\,.
\label{h-fun}
\end{equation}
The above line-element follows easily from the one in Eq. (\ref{rot-metric})
after projecting out the extra angular coordinates, by fixing their values,
and then setting $a=0$. As in the case of the projection of a Kerr-like
line-element on the brane, the induced metric still has an explicit
dependence on the number of extra dimensions that exist in the fundamental,
higher-dimensional theory.

The master equation (\ref{master3}) also takes now a much more simplified
form, namely
\begin{equation}
\label{radial1}
\Delta^s \frac{d}{dr}\left(\Delta^{1-s} \frac{d P_s}{dr}\right)+
\left[\frac{\omega^2 r^2}{h}+2i\omega s r-\frac{is\omega r^2 h'}{h}
-\Lambda\right] P_s (r)=0
\label{master-4}
\end{equation}
where now $\Lambda \equiv \tilde\Lambda_{sj}=j(j+1)-s(s-1)$. The above 
equation needs to be solved over the whole radial domain. We will first
review the results that follow from the analytic approach \cite{kmr1,kmr2}
and then move on to discuss the ones following from the numerical analysis
\cite{HK}. 

As mentioned in the previous subsection, the analytic treatment demands the
use of an approximate method in which the above equation is solved at the
near-horizon regime ($r \simeq r_H$) and far-field regime ($r \gg r_H$),
and the corresponding solutions are matched at an intermediate zone. 
In solving analytically Eq. (\ref{master-4}), the spin $s$ will remain an
arbitrary parameter of the differential equation; this will allow us to
solve the radial equation for particles with different spin $s$ in a
unified way.

Starting from the near-horizon (NH) regime, a change of variable 
$ r \rightarrow h(r)$ brings Eq. (\ref{master-4}) to the form \cite{kmr2}
\begin{eqnarray}
&~& \hspace*{-0.6cm} 
h\,(1-h)\,\frac{d^2 P_s}{dh^2} + \biggl[(1-s)\,(1-h) -\frac{(n+2s)}{(n+1)}
\,h\biggr]\,\frac{d P_s}{dh} + \nonumber \\[2mm]
&~& \hspace*{2cm} \biggl[\,\frac{(\omega r_H)^2}{(n+1)^2 h (1-h)} 
+ \frac{2is\,\omega r_H -\Lambda}{(n+1)^2 (1-h)}-
\frac{i s\,\omega r_H}{(n+1)\,h}\,\biggr] P_s=0\,.
\label{NH-1}
\end{eqnarray}
In the above, we have made use of the relation
\begin{equation}
\frac{dh}{dr}=\frac{(n+1)}{r}\,(1-h)\,,
\end{equation}
that follows from the definition (\ref{h-fun}). Making a further redefinition 
of the radial function, $P_s(h)=h^\alpha (1-h)^\beta F_s(h)$, the above equation
takes the form of a hypergeometric equation for the radial function $F_s$, i.e.
\begin{equation}
h\,(1-h)\,\frac{d^2 F_s}{dh^2} + [c-(1+a+b)\,h]\,\frac{d F_s}{dh} -ab\,F_s=0\,.
\label{hyper}
\end{equation}
The hypergeometric indices $(a,b,c)$ can be identified as
\begin{eqnarray}
a=\alpha + \beta +\frac{s + n\,(1-s)}{(n+1)}\,, \qquad 
b=\alpha + \beta\,, \qquad c=1-s + 2 \alpha\,.
\end{eqnarray}
The power coefficients $\alpha$ and $\beta$, in turn, are found by solving 
second-order algebraic equations leading to the results 
\begin{equation}
\alpha_{+} = s+\frac{i \omega r_H}{n+1}\,, \qquad
\alpha_{-} = -\frac{i \omega r_H}{n+1}\,, 
\end{equation}
and
\begin{equation}
\beta_{\pm} =\frac{1}{2 (n+1)}\,\biggl[\,1-2s \pm \sqrt{(1+2j)^2 -
4 \omega^2 r_H^2 -8 is \omega r_H}\,\,\biggr]\,,
\label{al-be}
\end{equation}
respectively. Then, by using the general solution \cite{AS} of the
hypergeometric equation (\ref{hyper}), together with the aforementioned
relation between $P_s(h)$ and $F_s(h)$, we obtain the following general
solution for the radial function $P_s(h)$ in the near-horizon regime:
\begin{eqnarray}
&& \hspace*{-1cm}P_{NH}(h)=A_- h^{\alpha}\,(1-h)^\beta\,F(a,b,c;h)
\nonumber \\[1mm] && \hspace*{2cm} +\,
A_+\,h^{-\alpha}\,(1-h)^\beta\,F(a-c+1,b-c+1,2-c;h)\,,
\label{NH-gen}
\end{eqnarray}
\noindent
where $A_{\pm}$ are arbitrary constants. For simplicity, in the above, we
have dropped the spin index $s$ from the radial function $P$.

An important boundary condition must be imposed on the general solution
(\ref{NH-gen}) at the horizon of the black hole: since nothing can escape
from the black hole, in the limit $r \rightarrow r_H$ the above general
solution must contain only incoming modes. In order to impose this
condition, we expand Eq. (\ref{NH-gen}) in the limit $r \rightarrow r_H$,
or equivalently $h \rightarrow 0$, and we obtain  
\begin{equation}
P_{NH} \simeq A_-\,h^\alpha + A_+\,h^{-\alpha} = 
A_-\,h^s\,\exp\Bigl(i \omega r_H^{n+2} y\Bigr) + 
A_+\,h^{-s}\,\exp\Bigl(-i \omega r_H^{n+2} y\Bigr)
\label{NH-alpha+}
\end{equation}
for $\alpha=\alpha_+$, and 
\begin{equation}
P_{NH} \simeq 
A_-\,\exp\Bigl(-i \omega r_H^{n+2} y\Bigr) + 
A_+\,\exp\Bigl(i \omega r_H^{n+2} y
\Bigr) \label{NH-alpha-}
\end{equation}
for $\alpha=\alpha_-$. In the above, we have used the `tortoise' coordinate
$y$ defined in Eq. (\ref{tortoise}). The aforementioned boundary condition 
at the horizon demands that $A_{-}=0$, in the former case, and $A_{+}=0$, in
the latter case. However, the remaining term in the asymptotic solution for
$\alpha=\alpha_+$ then describes an incoming wave with a diverging amplitude
at the horizon. This is clearly an irregular solution, therefore the choice
$\alpha=\alpha_+$ must be discarded altogether. On the other hand, the
choice $\alpha=\alpha_-$ leads to a regular incoming wave with
amplitude unity at the horizon. Finally, we need to make a choice for the
sign appearing in the expression of the $\beta$ coefficient. One can easily
see that the criterion for the convergence of the hypergeometric function
$F(a,b,c;h)$, i.e. ${\bf Re}\,(c-a-b)>0$, clearly demands that we choose
$\beta=\beta_{-}$.

Having found the solution in the near-horizon regime, we now turn our 
attention to the far-field asymptotic regime. In the limit $r \gg r_H$,
or $h \rightarrow 1$, Eq. (\ref{master-4}) assumes a simplified
form, which reads \cite{kmr2}
\begin{equation}
\frac{d^2 P}{d r^2} + \frac{2 (1-s)}{r}\,\frac{d P}{d r} +
\biggl(\omega^2 +\frac{2is \omega}{r} -\frac{\Lambda}{r^2} \biggr)\,P=0\,.
\end{equation}
Another redefinition of the radial function, $P=e^{-i \omega r} \,r^{j+s}\,
\tilde P(r)$, and a change of variable, $z=2 i \omega r$, puts the above
equation in the form of a confluent hypergeometric equation
\begin{equation}
z\,\frac{d^2 \tilde P}{d z^2} + (b- z)\,\frac{d \tilde P}{d z} - a \tilde P=0\,,
\end{equation}
with $a=j-s+1$ and $b=2j+2$. The general solution of the above differential
equation is given by \cite{AS}
\begin{equation}
\tilde P(z)=B_+\,M(a,b,z) + B_-\,U(a, b, z)\,,
\end{equation}
where $M$ and $U$ are the Kummer functions, and $B_\pm$ are arbitrary
coefficients. Then, the complete solution for the radial function $P(r)$
at $r \gg r_H$ can be written as
\begin{eqnarray}
&& \hspace*{-0.5cm} P_{FF}(r) = e^{-i \omega r} \,r^{j+s}\,\Bigl[B_+
\,M(j-s+1,\,2j+2,\,2i\omega r) \nonumber \\
&& \hspace*{4.5cm}+ B_-\,U(j-s+1,\,2j+2,\,2 i \omega r)\Bigr]\,.
\label{FF-sol}
\end{eqnarray}

In order to construct a complete solution for the radial function $P(r)$,
we need to match the two asymptotic solutions at an intermediate zone.
Although they look different, we will see that they actually take a very
similar form in the intermediate regime, which facilitates considerably
their matching. In order to see that, we first shift the hypergeometric
function towards large values of $r$. This can be done by using a standard
linear transformation formula \cite{kmr2,AS}, that changes its argument
from $h$ to $1-h$. Then, expanding in the limit $r \rightarrow \infty$,
or $h \rightarrow 1$, we obtain
\footnote{In order to obtain the simple powers of $r$ shown in the
expression below, the low-energy limit, $\omega r_H \ll 1$, has been
taken in the expression of the $\beta$ coefficient in the first term of
Eq. (\ref{NH-gen}). No expansion has been made in the arguments of the
Gamma functions.}
\begin{eqnarray}
P_{NH}(h) &\simeq& A_-\,\Bigl(\frac{r}{r_H}\Bigr)^{j+s}\,
\frac{\Gamma(1-s+2\alpha)\,
\Gamma\Bigl(-2\beta+\frac{1-2s}{n+1}\Bigr)}{\Gamma(\alpha-\beta+1-s)\,
\Gamma\Bigl(\alpha-\beta+\frac{1-2s}{n+1}\Bigr)}\nonumber \\[1mm]
&+& A_-\,\Bigl(\frac{r_H}{r}\Bigr)^{j-s+1}
\,\frac{\Gamma(1-s+2\alpha)\,\Gamma\Bigl(2\beta-\frac{1-2s}{n+1}\Bigr)}
{\Gamma(\alpha+\beta)\,\Gamma\Bigl(\alpha+\beta+\frac{s+n\,(1-s)}{n+1}\Bigr)}\,.
\label{NH-large}
\end{eqnarray}
We then expand the far-field solution (\ref{FF-sol}) in the limit
$\omega r \ll 1$, and we take
\begin{equation}
P_{FF}(r)= B_+\,r^{j+s} + \frac{B_-}{r^{j-s+1}}\,
\frac{\Gamma(2j+1)}{\Gamma(j-s+1)\,(2 i \omega)^{2j+1}}\,.
\label{FF-small}
\end{equation}
The matching of the two solutions (\ref{NH-large}) and (\ref{FF-small})
clearly provides relations between the integration constants appearing
in the expressions of the asymptotic solutions. These read
\begin{eqnarray}
B_+ &=& \frac{A_-}{r_H^{j+s}}\,\frac{\Gamma(1-s+2\alpha)\,
\Gamma\Bigl(-2\beta+\frac{1-2s}{n+1}\Bigr)}{\Gamma(\alpha-\beta+1-s)\,
\Gamma\Bigl(\alpha-\beta+\frac{1-2s}{n+1}\Bigr)}\,,\label{B+}\\[2mm]
B_- &=& \frac{A_-\,r_H^{j-s+1}\,(2 i \omega)^{2j+1}\,\Gamma(1-s+2\alpha)\,
\Gamma\Bigl(2\beta-\frac{1-2s}{n+1}\Bigr)\,\Gamma(j-s+1)}
{\Gamma(\alpha+\beta)\,\Gamma\Bigl(\alpha+\beta+\frac{s+n\,(1-s)}{n+1}\Bigr)
\,\Gamma(2j+1)}\,. \label{B-}
\end{eqnarray}
The above relations complete the determination of the solution for the radial
function $P_s(r)$ that describes the propagation of scalars, fermions and
gauge-bosons in the background of a projected $(4+n)$-dimensional
Schwarzschild-like black hole on a 4-dimensional brane.

We may now proceed to the calculation of the greybody factors for each type
of field. For this, we need the amplitudes of the  incoming and outgoing
modes at infinity. This can be found by taking the limit $r \rightarrow
\infty$ of the far-field solution (\ref{FF-sol}). Then, we obtain
\begin{equation}
P_s^{(\infty)}(r) = A^{(\infty)}_{in}\,\frac{e^{-i \omega r}}
{(2 \omega r)^{1-2s}} + A^{(\infty)}_{out}\,\frac{e^{i \omega r}}
{(2 \omega r)} + ...
\label{sol-inf}
\end{equation}
with
\begin{eqnarray}
A^{(\infty)}_{in} &=& \frac{e^{-i\pi(j-s+1)/2}}{(2\omega)^{j+s}}\,
\biggr[B_- + \frac{B_+\,e^{i\pi(j-s+1)}\,\Gamma(2j +2)}
{\Gamma(j+s+1)}\biggr]\,,\label{Ain}\\[3mm]
A^{(\infty)}_{out} &=&
\frac{B_+\,e^{-i\pi(j+s+1)/2}\,\Gamma(2j+2)}{\Gamma(j-s+1)\,
(2\omega)^{j+s}}\,. \label{Aout}
\end{eqnarray}

A few comments are here in order: for $s=0$, the above expression describes
the incoming and outgoing mode of a scalar field at infinity, both scaling
as $1/r$; this expression, therefore, contains the complete information for
the propagating scalar field and the two amplitudes, $A^{(\infty)}_{in}$
and $A^{(\infty)}_{out}$, can be used to determine the reflection coefficient
${\cal R}$, and from that the absorption coefficient ${\cal A}$, as follows
\begin{equation}
|{\cal A}^{(0)}_j|^2=1-|{\cal R}^{(0)}_j|^2= 
1-\Biggl|\frac{A_{out}^{(\infty)}}{A_{in}^{(\infty)}}\Biggr|^2\,.
\label{scalars}
\end{equation}
On the other
hand, for $s \neq 0$ and $s=+|s|$, Eq. (\ref{sol-inf}) describes a dominant
incoming mode and a suppressed outgoing one; vice versa, for $s=-|s|$, 
the outgoing mode dominates while the incoming one is greatly suppressed.
This is merely a reflection of the fact that, for $s \neq 0$, the
propagating field has more than one components, each one of which mainly
describes either the incoming mode or the outgoing mode. What is fortunate
is that we do not need to know the complete solution at infinity as long
as we can compute the total incoming flux ${\cal F}_{in}$ at the horizon
and at infinity. Then, the absorption coefficient may be directly determined
through the following expression 
\begin{equation}
|{\cal A}^{(s)}_j|^2= \frac{{\cal F}_{in}^{\,(h)}}
{{\cal F}_{in}^{\,(\infty)}}\,.
\label{absorption}
\end{equation}
The calculation of the incoming flux is indeed possible for the
spherically-symmetric Schwarzschild-like background that we consider here
\footnote{Note that for more complicated gravitational backgrounds, the
calculation of the flux can be highly non-trivial, especially close to
the horizon. In that case, the complete solution (both incoming and outgoing
modes) at infinity must be determined.}, so the only component whose
expression we need to calculate, for fields with a non-zero spin, is the
upper one, with $s=+|s|$. 

The incoming flux of a fermionic field can be computed from the radial
component of the conserved current, $J^\mu=\sqrt{2}\,\sigma^\mu_{AB}\,
\psi^A \bar\psi^B$, integrated over a two-dimensional sphere, first at
infinity and then at the horizon \cite{CL2,kmr2}. For gauge bosons, the 
($tr$)-component of the energy momentum tensor $T^{\mu\nu}=2 \sigma^\mu_{AA'}
\sigma^\nu_{BB'} \psi^{AB} \bar\psi^{A'B'}$ can be used instead. The same
method for the calculation of the absorption coefficient can be equally well
applied also in the case of a scalar field with the use of the radial
component of the conserved current $J^\mu=h r^2\,(\psi \partial^\mu \psi^*-
\psi^* \partial^\mu \psi)$ - this method is completely
equivalent to the one described above, that makes use of the expression
of the reflection coefficient, while it allows us to write down a unified
expression for the absorption probability for fields with arbitrary spin.
Substituting the results for the incoming fluxes in Eq. (\ref{absorption}),
we may finally write
\begin{equation}
|{\cal A}^{(s)}_j|^2=(2 \omega r_H)^{2(1-2s)}
\biggl|\frac{A^{(h)}_{in}}{A^{(\infty)}_{in}}\biggl|^2\,,
\label{unified}
\end{equation}
where $A^{(h)}_{in}=A_-$, and $A^{(\infty)}_{in}$ is given in Eq. (\ref{Ain}). 
In the above, we have taken into account the fact that, for fermions and gauge
fields, the lower component contributes little to the in-falling flux both at
infinity and at the horizon. We should note here that the coefficient in
front of the ratio of the two amplitudes strongly depends on the field
normalization (\ref{sol-inf}) at infinity - although this is merely a
convention, care should be taken so that this coefficient correctly reflects
the chosen normalization~\footnote{For this reason, different coefficients
appear in the expressions of the absorption probabilities
$|{\cal A}^{(s)}_j|^2$ in Refs. \refcite{kmr2} and \refcite{HK}, where
different normalizations for the field at infinity were used.}.

Substituting the expression of $A^{(\infty)}_{in}$ in Eq. (\ref{unified})
and using the relations (\ref{B+}) and (\ref{B-}), we obtain (for more
information on the mathematical details omitted below, see 
Ref. \refcite{kmr2})
\begin{equation}
|{\cal A}^{(s)}_j|^2 = \frac{(2\omega r_H)^{2j+2-2s}}
{|\Gamma(1-s+2 \alpha)|^2\, |\,C\,(\omega r_H)^{2j+1} + D\,|^2}\,,
\label{total} 
\end{equation}
where the coefficients $C$ and $D$ stand for
\begin{eqnarray}
C &=& \frac{2^{2j+1}\,e^{i\pi\,(s-1/2)}\,\Gamma\Bigl(2\beta-\frac{1-2s}{n+1}
\Bigr)\,\Gamma(j-s+1)}{\Gamma(\alpha+\beta)\,\Gamma\Bigl(\alpha+\beta+
\frac{s+n\,(1-s)}{n+1}\Bigr)\,\Gamma(2j+1)}\,, \label{C}\\[2mm]
D &=& \frac{\Gamma(2j+2)\,\Gamma\Bigl(-2\beta+\frac{1-2s}{n+1}\Bigr)}
{\Gamma(\alpha-\beta+1-s)\,\Gamma\Bigl(\alpha-\beta+\frac{1-2s}{n+1}\Bigr)
\,\Gamma(j+s+1)}\,.\label{D}
\end{eqnarray}

Since we study the emission of brane-localized modes, this is clearly a
four-dimensional process, therefore the relation (\ref{grey-n}) [or
(\ref{greyb})] between the greybody factor and the absorption probability
takes the simplified form
\begin{equation}
\sigma^{(s)}_{j,n}(\omega) = \frac{\pi}
{ \omega^{2}}\,(2j+1)\,|{\cal A}^{(s)}_j|^2=\frac{A_H}
{ (2\omega r_H)^{2}}\,(2j+1)\,|{\cal A}^{(s)}_j|^2\,,
\end{equation}
where $A_H=4 \pi r_H^2$, or finally
\begin{equation}
\sigma^{(s)}_{j,n}(\omega) = \frac{(2\omega r_H)^{2j-2s} \,(2j+1)\,A_H}
{|\Gamma(1-s+2 \alpha)|^2\, |\,C\,(\omega r_H)^{2j+1} + D\,|^2}\,.
\label{final}
\end{equation}
The coefficients $C$ and $D$ in the above expression depend, through Gamma
functions, on the parameters of the theory, therefore the dependence of
the greybody factor on the spin $s$, the energy $\omega$, or the number of
extra dimensions $n$ is not clear. Nevertheless, the above expression can
be easily plotted and thus reveal the desired dependence. Before however
discussing those results, we might attempt to derive some simple, elegant
expressions for the greybody factor by taking the low-energy limit
$\omega r_H \ll 1$. In this limit, we may express each one of the
coefficients $C$ and $D$ as a power series in $(\omega r_H)$ and keep
only the leading term.
Then, we may see \cite{kmr2} that the term proportional to $C$,
in the denominator of Eq. (\ref{final}), is sub-dominant compared to the 
one proportional to $D$ -- that comes out to be independent of $(\omega r_H)$
at leading order -- and thus the former can be ignored. Depending on the
spin of the particle, additional powers of $(\omega r_H)$ may come from the
measure of the remaining Gamma function in the denominator of Eq.
(\ref{final}). While for  scalars and fermions, the zeroth-order term is
$(\omega r_H)$-independent, i.e. 
\begin{eqnarray}
\frac{1}{|\Gamma(1+2 \alpha)|^2} &=& 1 + \frac{2 \pi^2 (\omega r_H)^2}
{3 (n+1)^2} + {\cal O}(\omega r_H)^4\,, \\[2mm]
\frac{1}{|\Gamma(1/2+2 \alpha)|^2} &=& \frac{1}{\pi} + \frac{2\pi\,(\omega r_H)^2}
{(n+1)^2} + {\cal O}(\omega r_H)^4\,,
\end{eqnarray}
for gauge bosons this is not the case and extra powers of $(\omega r_H)$
come up 
\begin{equation}
\frac{1}{\Gamma(2\alpha)\,\Gamma(-2\alpha)}=\frac{4\,(\omega r_H)^2}
{(n+1)^2} + {\cal O}(\omega r_H)^4\,.
\end{equation}
Putting everything together, we obtain the following simplified, low-energy
expressions for the greybody factors
\begin{equation}
\sigma^{(0)}_{j,n}(\omega) = \frac{\pi\,(2j+1)}{(n+1)^2}\,
\frac{\Gamma(\frac{j+1}{n+1})^2\,\Gamma(1+\frac{j}{n+1})^2}
{\Gamma(\frac{1}{2}+j)^2\,\Gamma(1 +\frac{2j+1}{n+1})^2}\,
\biggl(\frac{\omega r_H}{2}\biggl)^{2j}\,A_H + ... \,,
\label{simple}
\end{equation}
for scalars, 
\begin{equation}
\sigma^{(1/2)}_{j,n}(\omega) = \frac{\pi\,(2j+1)\,2^{-(4j+2)/(n+1)}}
{4\,\Gamma(j+1)^2}\,\biggl(\frac{\omega r_H}{2}\biggl)^{2j-1}\,A_H + ...\,,
\label{fermionsi}
\end{equation}
fermions, 
\begin{equation}
\sigma^{(1)}_{j,n}(\omega) =\frac{(2j+1)}{(n+1)^2}\,
\left(\frac{\Gamma\Bigl(\frac{j}{n+1}\Bigr)\,
\Gamma\Bigl(\frac{j+1}{n+1}\Bigr)\,\Gamma(j+2)}
{\Gamma\Bigl(\frac{2j+1}{n+1}\Bigr)\,\Gamma(2j+2)}\right)^2\,
(2\omega r_H)^{2j}\,A_H + ...\,,
\label{gaugesi}
\end{equation}
and gauge bosons. By choosing different values for $j \geq s$, the above
approximate expressions reveal that, at the low-energy regime, the greybody
factor for all types of fields gets suppressed as the angular momentum
number increases; on the other hand, the same expressions predict,
independently of the value of the spin $s$, an enhancement of the value of
the greybody factor as the number of extra dimensions, that are projected
on the brane, increases.

\begin{figure}[t]
\begin{center}
\begin{tabular}{c} \hspace*{-0.3cm}
\epsfig{file=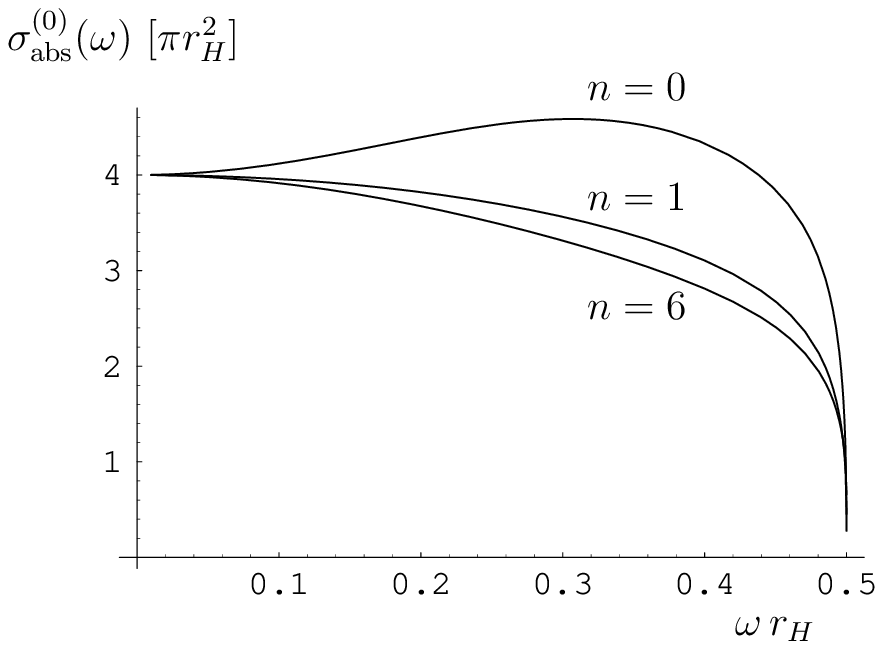,width=4.1cm} 
\epsfig{file=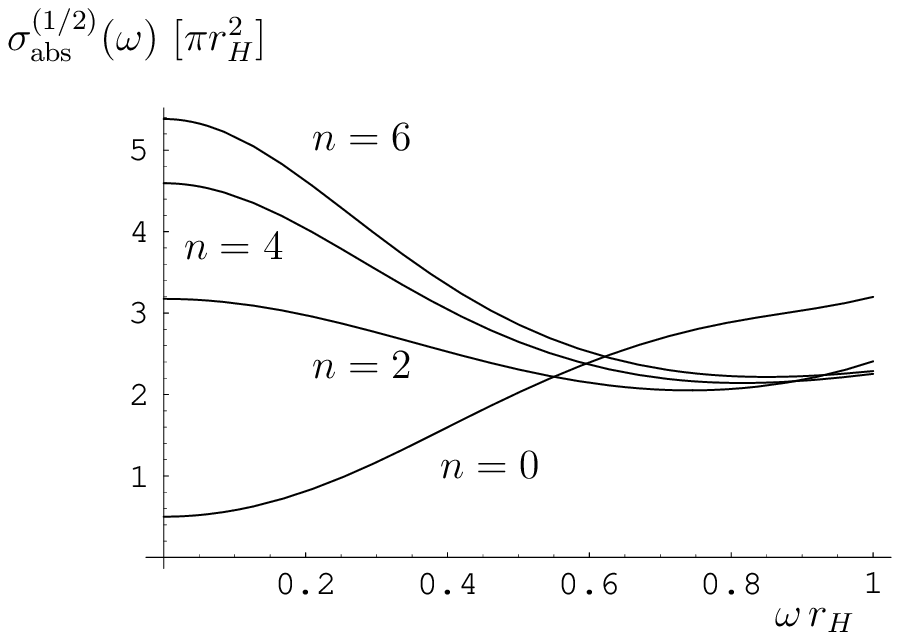,width=4.2cm} \hspace*{0.1cm}
\epsfig{file=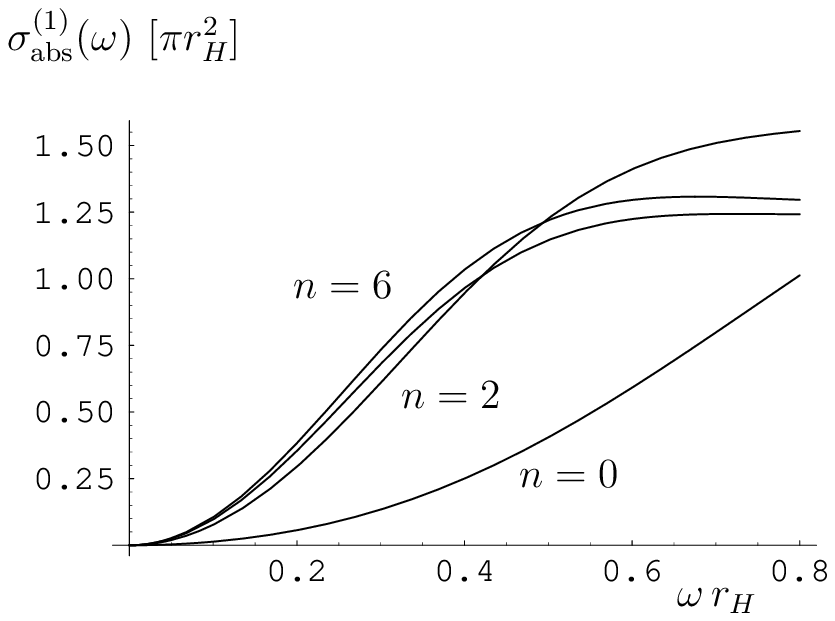,width=4.1cm} \end{tabular}
\end{center}
\caption{Analytical results for the greybody factors for emission of
{\bf (a)} scalars, {\bf (b)} fermions, and {\bf (c)} gauge bosons from
a $(4+n)$-dimensional black hole on the brane.}
\end{figure}

Let us note at this point that the {\it full analytic} expression
Eq. (\ref{final}) has been derived in the low-energy approximation:
the low-energy limit was taken in the expression of the $\beta$ coefficient,
during the matching of the two asymptotic solutions. The {\it simplified
analytic} expressions, Eqs. (\ref{simple})-(\ref{gaugesi}), have been
derived in the same low-energy limit by further expanding Eq. (\ref{final}).
However, it is clear that the more we expand an expression, the more
we limit its validity, as more and more terms are left out. The
simplified expressions (\ref{simple})-(\ref{gaugesi}), therefore, are valid
in a more narrow low-energy regime than the full analytic one (\ref{final}).
Moreover, the information that we obtain from the simplified expressions
is trustable only as long as it is deduced from the lowest partial wave, 
$j=|s|$, which always gives the main contribution to the greybody factor. 
The aforementioned dependence of $\sigma^{(s)}_{j,n}(\omega)$ on $j$ for
all types of fields, as well as the dependence on $n$ for fermions and
gauge bosons were indeed derived by looking at the leading term of the
lowest partial wave. However, in the case of scalar fields the dominant 
partial wave ($j=0$) is found to be independent of $n$ having the
form $\sigma^{(0)}_{0,n}(\omega) \simeq 4\pi r_H^2 =A_H$, which is
a well-known property of the low-energy behaviour of the greybody factor
for scalar fields. One then is forced to look at higher-order terms for
the desired dependence on $n$. Because the higher partial waves are
enhanced with $n$, one may naively conclude that the greybody factor itself
increases. However, it turns out that the next-to-leading order terms in
the expansion of the lowest partial-wave, denoted by ellipses in
Eq. (\ref{simple}), are of the same order as the leading terms of the higher
partial waves, and thus should also be taken into account. When this is done,
we are finally led to a decrease of the greybody factor for scalar fields
with $n$. An extreme caution is therefore necessary when one deals with
simplified expressions as the Eqs. (\ref{simple})-(\ref{gaugesi}).

On the other hand, for more accurate, and thus more informative, evaluations
of the greybody factors and associated emission rates, we clearly need to
use the full analytic result (\ref{final}). Although an approximate result
itself, this expression leads to a behaviour for the greybody factors that
is in excellent agreement with the exact behaviour in the low-energy regime,
and in fairly good agreement with the one in the intermediate-energy regime.
Plotting Eq. (\ref{final})
as a function of the energy parameter $\omega r_H$, for different values
of $s$, leads to the behaviour depicted in Figures 1(a,b,c) \cite{kmr2}, where 
$\sigma_{\rm abs}^{(s)}$ is the greybody factor (\ref{final}) summed over $j$
up to the third partial wave -- it is easy to see that any higher partial
wave has a negligible contribution to the final result in this energy regime.
As mentioned above, the greybody factor for scalar fields is found \cite{kmr2}
to decrease with $n$, while the ones for fermions and gauge bosons are
enhanced as $n$ increases, at least up to intermediate energies.
While scalars and fermions have a non-vanishing greybody factor as 
$\omega r_H \rightarrow 0$, the one for the gauge bosons vanishes.

The full analytic result for the greybody factor (\ref{final}) may then
be substituted in the expression of the power flux for emission in the
four-dimensional spacetime, namely
\begin{equation}
\frac{dE^{(s)}(\omega)}{dt} = \sum_{j} \sigma_{j,n}^{(s)}(\omega)\,
{\omega^3  \over \exp\left(\omega/T_{H}\right) \pm 1}\,
\frac{d \omega}{2\pi^2}\,.
\label{emission}
\end{equation}
Figures 2(a,b,c) depict the behaviour of the energy emission rates for
particles with spin 0, $\frac{1}{2}$ and 1 in the low- and intermediate-energy
regime. Despite the different low-energy behaviour of the greybody factors
for scalars, fermions and gauge bosons, depicted in Figure 1, the 
corresponding power emission rates, as well as the flux emission rates,
exhibit a universal behaviour according to which the energy, and the
number of particles, emitted per unit time and energy interval is strongly
enhanced, as $n$ increases. This was to be anticipated: we remind the
reader that the temperature of the black hole is given by the relation
$T_H=(n+1)/4 \pi r_H$, therefore, for fixed $r_H$, the temperature of
the black hole increases as $n$ increases. This simply means that the energy
of the black hole available for the emission of particles also increases,
and this is reflected in the enhancement of the power and flux rates.
The effect of the greybody factor, on the other hand, is to suppress or
enhance -- depending on the spin of the particle, the dimensionality of
spacetime and the energy regime that we are looking at -- the emission rates
compared to the ones derived by using its high-energy, geometrical optics 
value. This suppression or enhancement is then reflected at the location
of the peak of the emission curve. For example, having seen that the
greybody factor for gauge bosons is very much suppressed at the low-energy
regime, compared to scalars and fermions, we expect that more spin-1 particles
will be emitted in the intermediate and high-energy regime, thus shifting
the peak of the gauge boson emission curve towards higher energies. This feature
will indeed be obvious when the complete spectra are derived.

\begin{figure}[t]
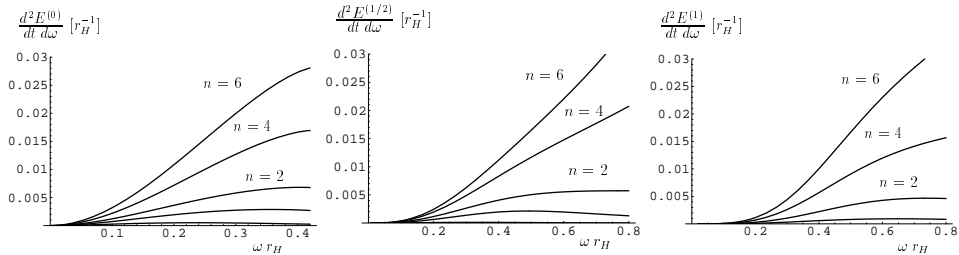

\begin{center}
\begin{tabular}{c} \hspace*{-0.3cm}
\epsfig{file=sc-dec-nolog.ps,width=4.1cm} 
\epsfig{file=fer-dec-nolog.ps,width=4.2cm}
\epsfig{file=gb-dec-nolog.ps,width=4.1cm} \end{tabular}
\end{center}
\caption{Analytical results for the energy emission rates for {\bf (a)}
scalars, {\bf (b)} fermions, and {\bf (c)} gauge bosons from a 
$(4+n)$-dimensional black hole on the brane.}
\end{figure}

Each one of the curves depicted in Figure 2 has a blackbody profile.
Nevertheless, the approximate analytic expression (\ref{final}) for the
greybody factor, and thus the corresponding emission rates, are accurate
only in the low-energy regime while at intermediate energies it mainly
provides qualitative agreement with the exact behaviour. Any attempt to
derive emission rates in the high energy regime by using this expression
is bound to lead to wrong results. For this reason, the graphs in Figure 2
extend only up to intermediate energies for fermions and gauge bosons,
while for scalars, our analytic formula breaks down much earlier [see
Fig. 1(a)].

Since the analytic results can accurately describe only the part of the 
``greybody" curve that extends over the low-energy regime, a numerical
analysis is clearly necessary in order to derive the complete radiation
spectrum. In that case, the master radial equation (\ref{master3}) can
be solved through numerical integration \cite{HK} under the same boundary
conditions that were used in the analytic approach: a purely incoming mode
at the horizon of the black hole is integrated outwards until the asymptotic
solution (\ref{FF-sol}) is reached. The determination of the asymptotic
coefficients, $A_-$ and $A^{(\infty)}_{in}$, can then be used in conjunction
with Eq. (\ref{unified}) to define the absorption coefficient, and from that
the greybody factors and emission rates \cite{HK}, as described above.
The numerical manipulation of this problem is not without technical
difficulties itself: as the spin of the particle increases, it becomes
more and more difficult to accurately determine the two asymptotic 
coefficients; employing various transformations of the radial equation
can help remedy this problem (for more information on this point, the
reader is referred to Ref. \refcite{HK}).
\begin{figure}[b]
\begin{center}
\begin{tabular}{c}
\epsfig{file=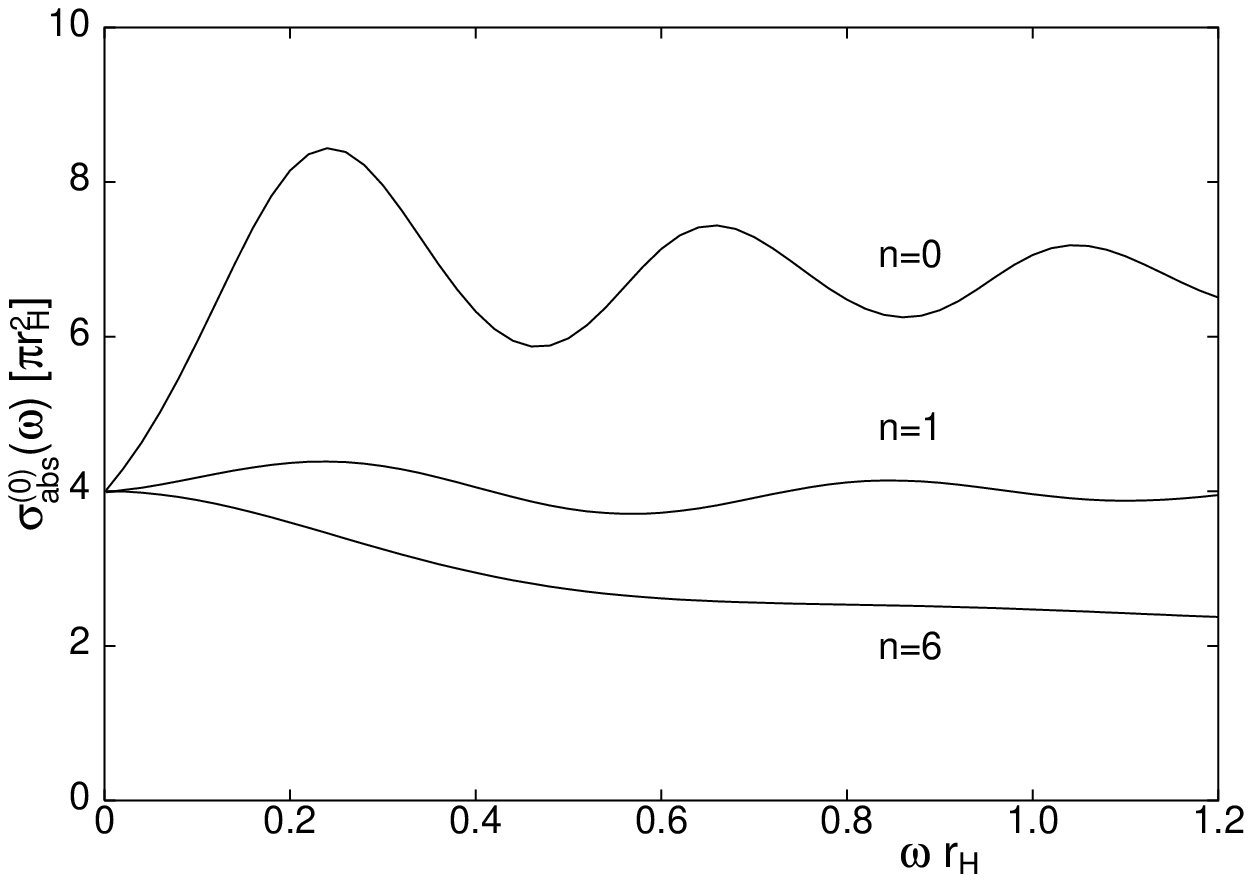,width=5.5cm} \hspace*{0.1cm}
\epsfig{file=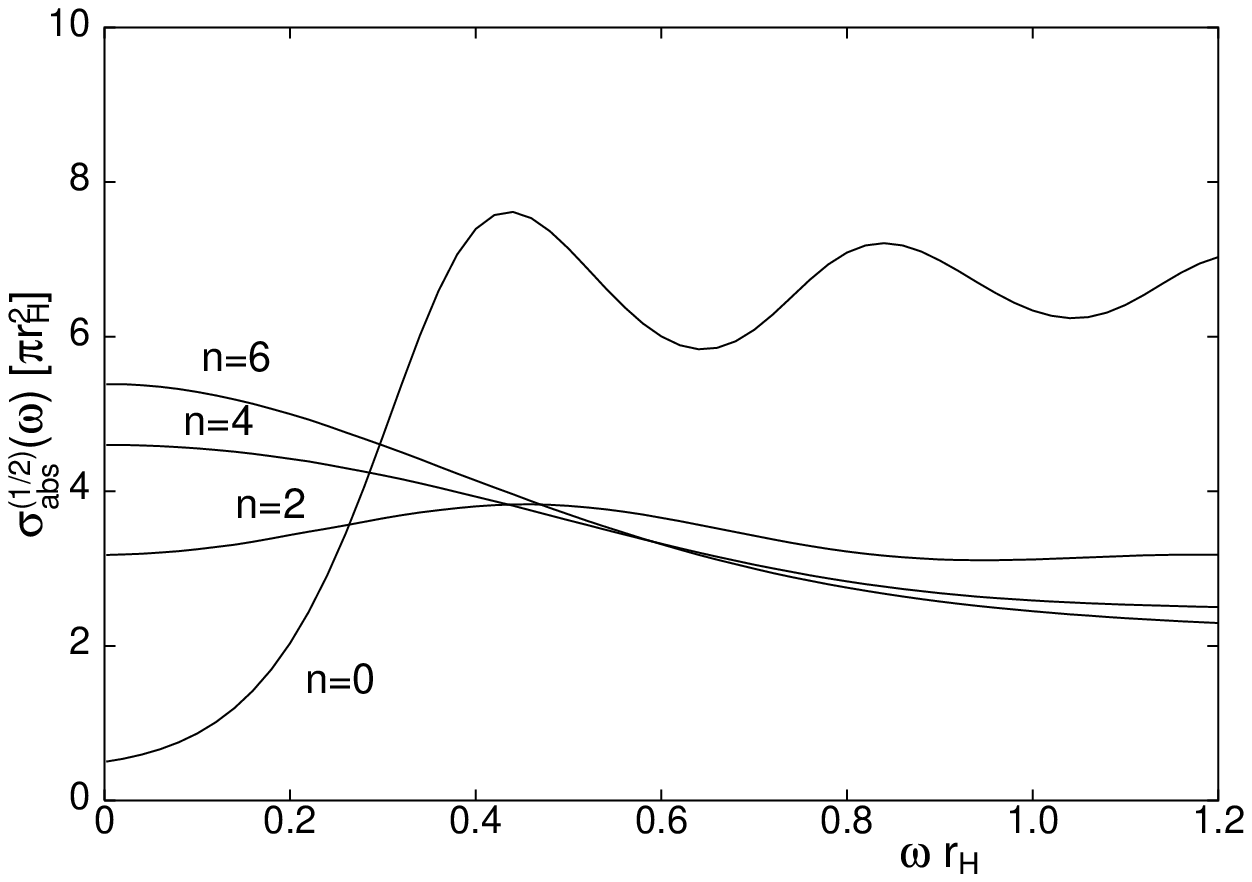,width=5.5cm}
\\[2mm] \epsfig{file=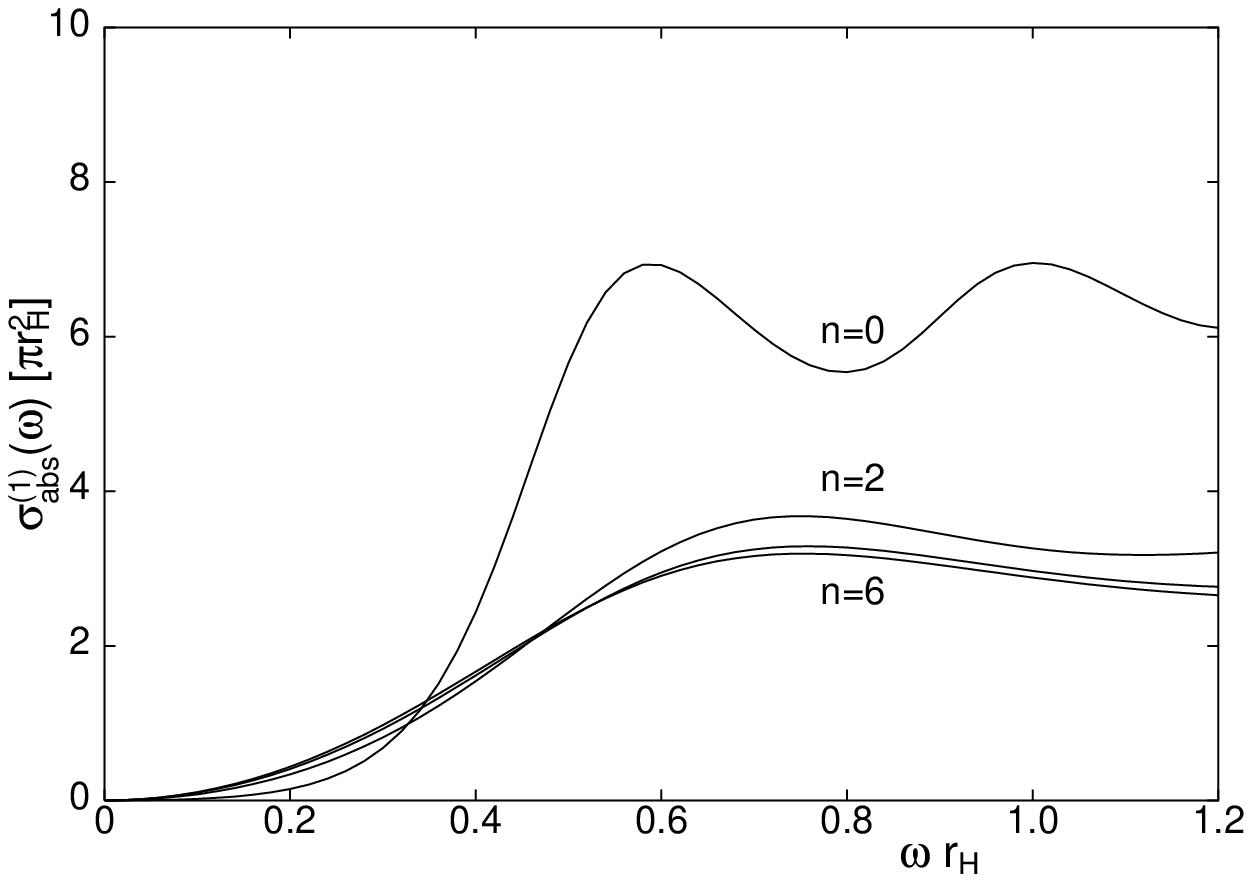,width=5.5cm} \end{tabular}
\end{center}
\caption{Numerical results for the greybody factors for emission of
{\bf (a)} scalars, {\bf (b)} fermions, and {\bf (c)} gauge bosons from
a $(4+n)$-dimensional black hole on the brane.}
\end{figure}

The numerical integration was performed in Ref. \refcite{HK}
where both greybody factors and emission rates for particles with spin 0,
$\frac{1}{2}$ and 1 were computed. The exact results for the greybody
factors are displayed in Figures 3(a,b,c). Their complex behaviour 
depending on the spin of the particle emitted, the number of the extra
dimensions projected on the brane and the energy regime studied, is evident. 
At the low-energy regime, the exact results closely follow the ones
derived by the analytic method. As $\omega r_H \rightarrow 0$, the greybody
factor for scalar fields reduces to $4\pi r_H^2$, thus revealing the
fact that the low-energy value of this quantity is indeed proportional 
to the area of the black hole horizon, even when a number of extra
dimensions is projected onto our four-dimensional spacetime. The
greybody factor for fermions also reduces to a constant value, different
for different values of $n$, while the one for gauge bosons goes to
zero as predicted by the analytic results. 

As the energy increases further,
the exact results start deviating from the analytic ones and they can be
seen adopting an oscillatory behaviour -- due to the late dominance of higher
partial waves -- around a high-energy asymptotic value. This asymptotic
value is the same  for all particle species, nevertheless it strongly
depends on the number of extra dimensions. In this regime, the greybody
factor assumes its geometrical optics limit value, already well-known
from the four-dimensional case \cite{page,sanchez,MTW,macG}. For a
massless particle in a circular orbit around a black hole, described by 
the line-element (\ref{metric-n}), its equation of motion,
$p^\mu p_\mu=0$, takes the form
\begin{equation}
\biggl(\frac{1}{r}\,\frac{dr}{d\varphi}\biggr)^2=\frac{1}{b^2}
-\frac{h(r)}{r^2}\,,
\end{equation}
where $b$ is the ratio of the angular momentum of the particle over
its linear momentum. The classically accessible regime is defined by
the relation $b< {\rm min}(r/\sqrt{h})$. Although the number of the
projected dimensions do not change the general structure of the above
equation, valid for motion in a four-dimensional, spherically-symmetric
background, the expression of the metric function $h(r)$ does change
as it carries an explicit dependence on $n$. Using the definition
(\ref{h-fun}), we can easily find that the closest the particle can get 
to the black hole is at a distance \cite{emparan}
\begin{equation}
b = r_c \equiv \biggl(\frac{n+3}{2}\biggr)^{1/n+1}\,
\sqrt{\frac{n+3}{n+1}}\,\,r_H\,.
\label{effective}
\end{equation}
The above expression correctly reproduces, for $n=0$, the four-dimensional
value of $r_c=3 \sqrt{3}\,r_H/2$ \cite{sanchez,MTW}. The radius $r_c$
defines the absorptive area of the black hole at high energies and, thus,
the corresponding value of the greybody factor, $\sigma_g = \pi r_c^2$,
which, being a constant, now describes a blackbody. Since $r_c$ decreases
as $n$ increases, the asymptotic greybody factor becomes more and more
suppressed as the number of extra dimensions projected onto the brane
gets larger. 

\begin{figure}[t]
\begin{center}
\begin{tabular}{c}
\epsfig{file=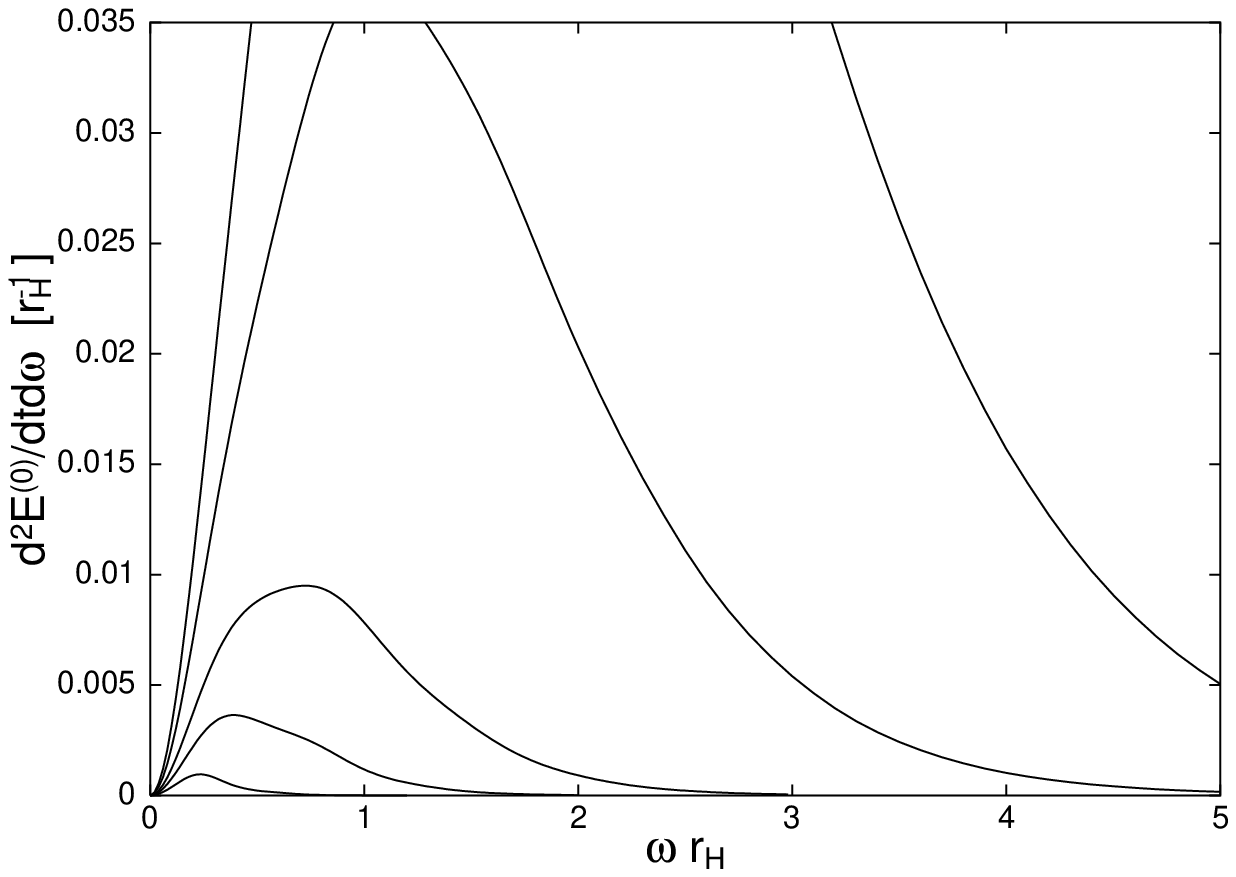,width=5.7cm} \hspace*{0.1cm}
\epsfig{file=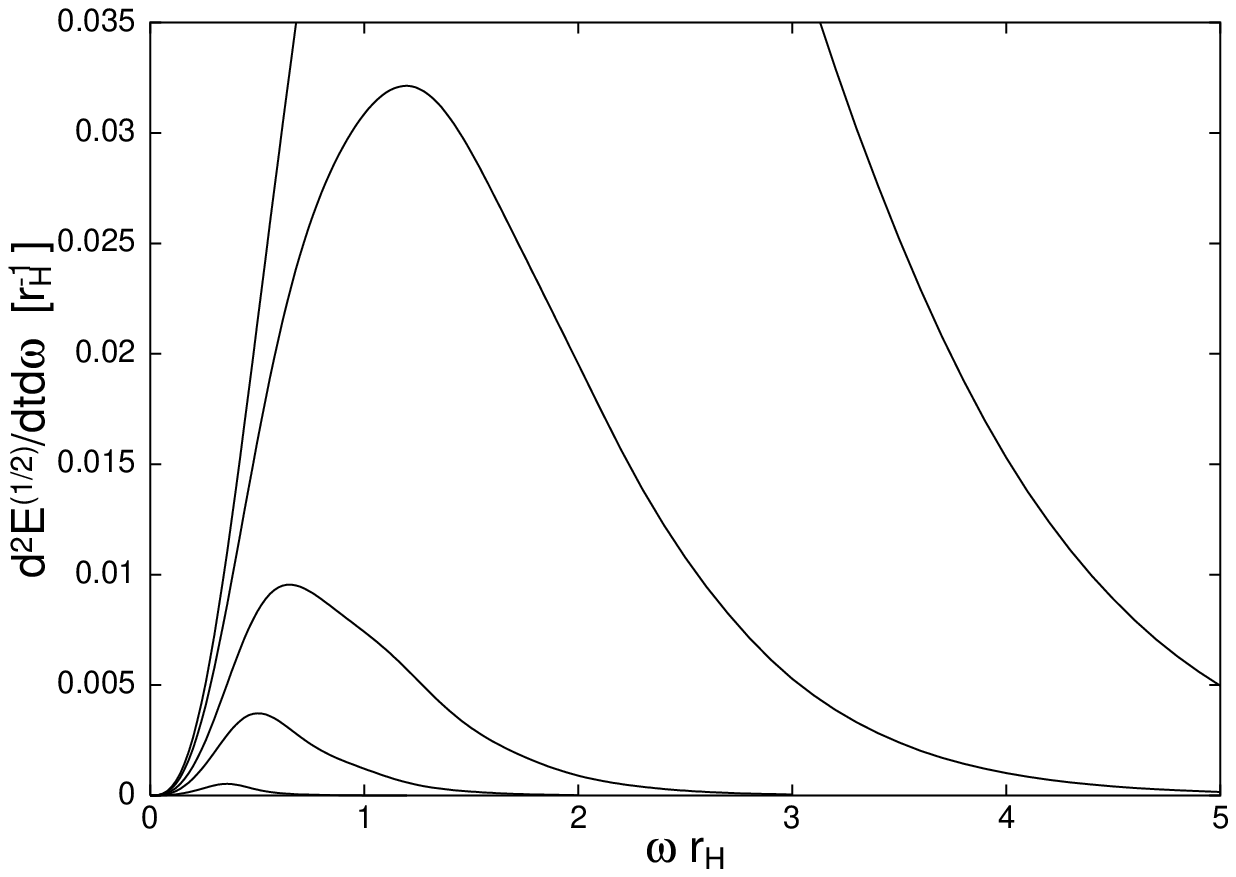,width=5.7cm}
\\[2mm] \epsfig{file=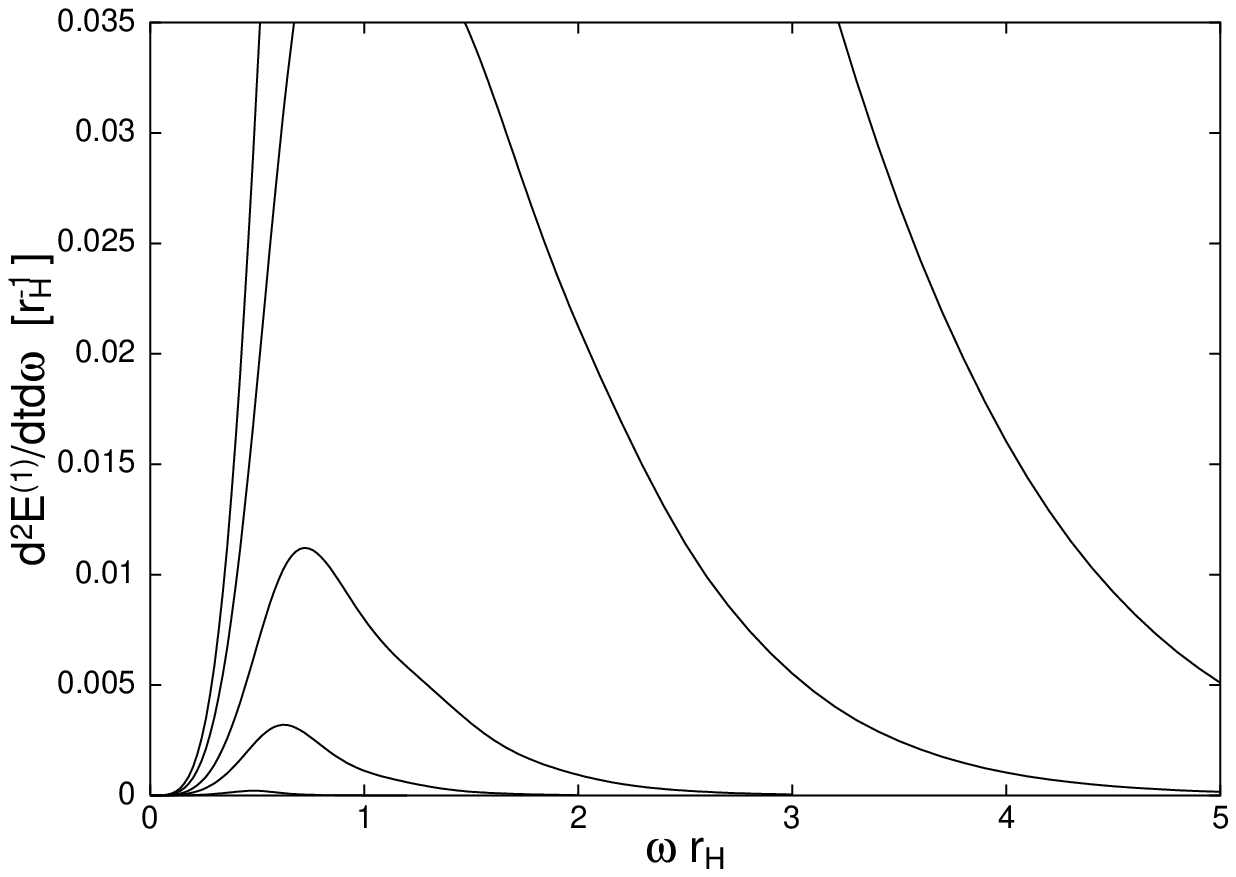,width=5.6cm} \end{tabular}
\end{center}
\caption{Numerical results for the energy emission rates for {\bf (a)}
scalars, {\bf (b)} fermions, and {\bf (c)} gauge bosons from a 
$(4+n)$-dimensional black hole on the brane.}
\end{figure}

Turning to the emission rates, the exact results for the energy emission
in the form of scalars, fermions and gauge bosons are depicted in
Figures 4(a,b,c) \cite{HK}. The complete emission curves have now been
constructed and can be easily compared for different values of $n$ and
for different species. For all types of particles, an important enhancement
of the energy emission rate takes place, as $n$ increases. This
was already predicted by the analytic results, however, the exact numerical
results, that are now available, allow us to accurately calculate the
enhancement of the emission rates. Integrating over the whole energy regime,
we obtain the total energy emissivities, which are displayed in Table 4
\cite{HK} normalized to the corresponding emissivities for $n=0$.
It is worth pointing out that the amount of energy emitted by the black hole
in the case of $n=7$ is 3 orders of magnitude larger than the one for $n=0$,
for all types of particles. A similar enhancement is observed in the number
of particles emitted by the black hole.

\begin{table}[h]
\tbl{Total energy emissivities for different values of $n$}
{\begin{tabular}{@{}ccccccccc@{}} \toprule
$n$ & 0 & 1 & 2 & 3 & 4 & 5 & 6 & 7 \\ \colrule
\hspace*{1mm} Scalars  & 1.0 & 8.94 & 36.0 & 99.8 & 222 &
429 & 749 & 1220\\ 
\hspace*{1mm} Fermions  & 1.0 & 14.2 & 59.5 & 162 & 352 &
664 & 1140 & 1830\\ 
\hspace*{1mm} Gauge Bosons  & 1.0 & 27.1 & 144 & 441 & 1020 &
2000 & 3530 & 5740
\\ \botrule
\end{tabular}}
\end{table}

Another important question, that one can pose, is what type of particles
the black hole `prefers' to emit, as the number of extra dimensions $n$
changes. From the four-dimensional analyses \cite{page,macG} we know that,
for $n=0$, most of the energy of the black hole is emitted in the form of
scalar particles; this becomes clear by comparing the areas under the
emission curves, for particles with spin 0, $\frac{1}{2}$ and 1, in 
Figure 5(a). The emission curves in Figure 5(b) have been drawn for $n=2$:
the emission curves for scalars, fermions and gauge bosons now seem to 
define comparable areas. Increasing further the value of $n$,
i.e. $n=6$, we obtain the emission curves
shown in Figure 5(c). The situation has radically changed: the gauge
bosons are now the dominant `channel' for the emission of the black hole
energy, with the scalars and fermions coming second and third, respectively.
Accurate estimates for the relative emissivities have been made \cite{HK},
and are displayed in Table 5, normalized to the emissivity of scalar fields.
The 3-species ratios shown below can equally well be used as signatures for
the dimensionality of spacetime if the emitted spectrum of Hawking radiation
is successfully observed.

\begin{figure}[t]
\begin{center}
\begin{tabular}{c}
\epsfig{file=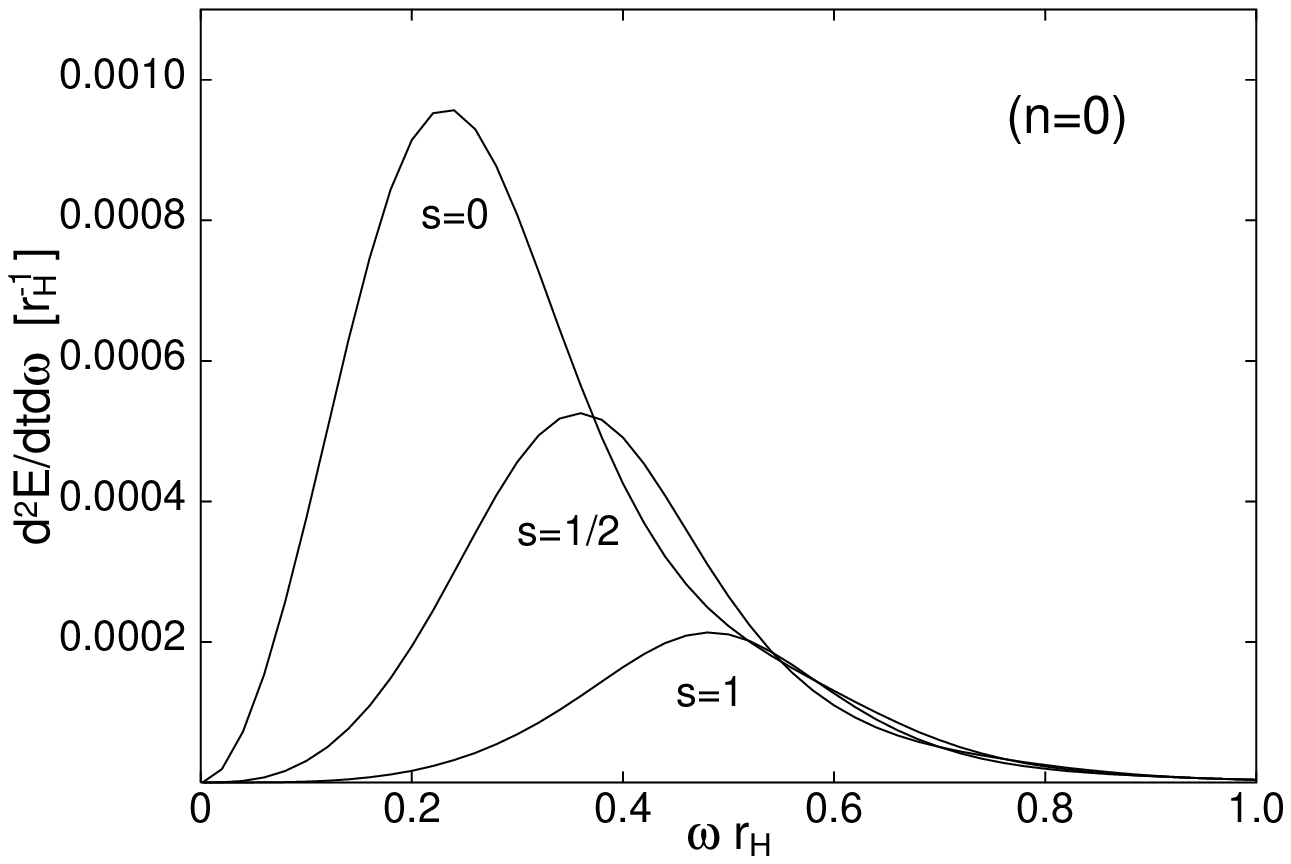,width=4cm} \hspace*{0.1cm}
\epsfig{file=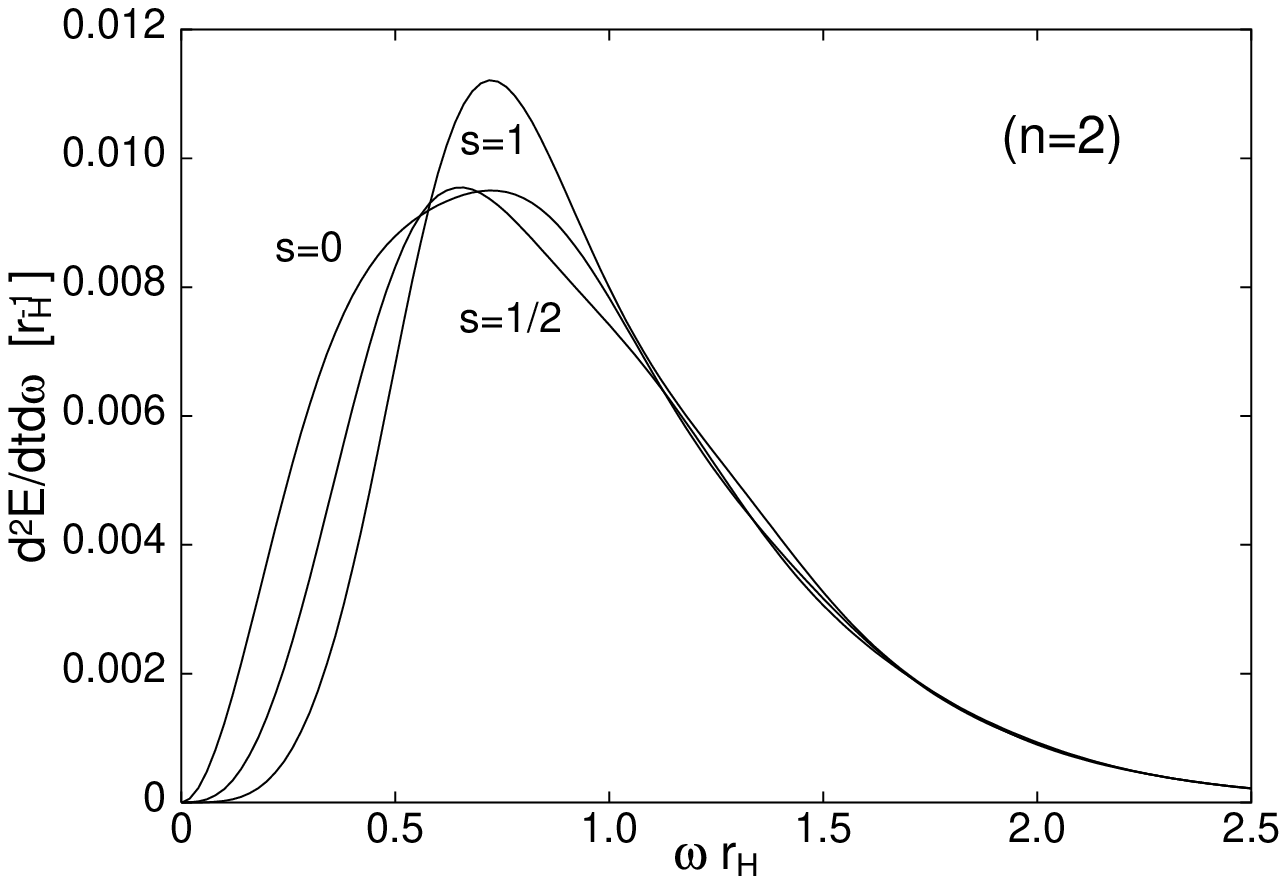,width=4cm}
\epsfig{file=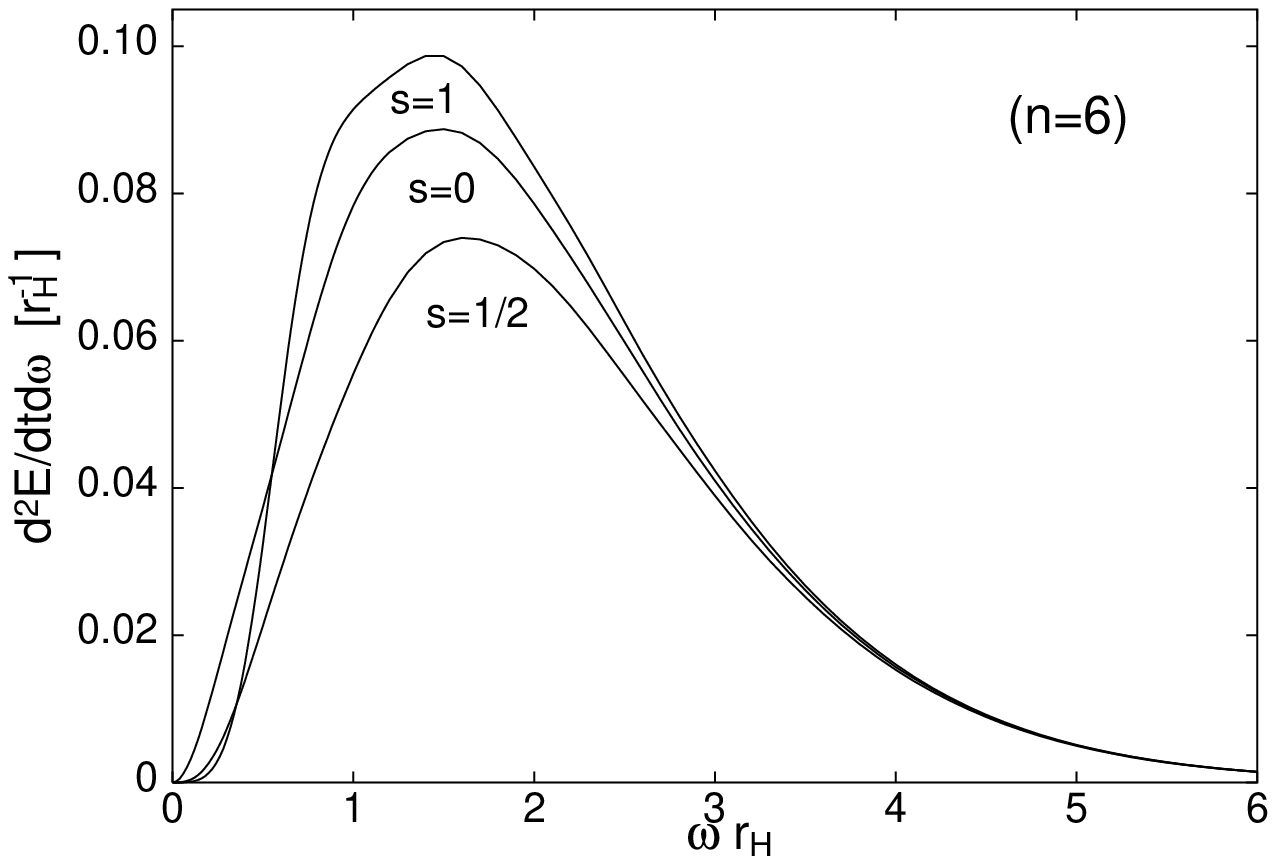,width=4cm} \end{tabular}
\end{center}
\caption{Energy emission rates for scalars, fermions and gauge bosons
for {\bf (a)} $n=0$, {\bf (b)} $n=2$, and {\bf (c)} $n=6$.}
\end{figure}

\begin{table}[h]
\tbl{Relative energy emissivities for different values of $n$}
{\begin{tabular}{@{}ccccccccc@{}} \toprule
$n$ & 0 & 1 & 2 & 3 & 4 & 5 & 6 & 7 \\ \colrule
\hspace*{1mm} Scalars  & 1.00 & 1.00 & 1.00 & 1.00 & 1.00 &
1.00 & 1.00 & 1.00\\ 
\hspace*{1mm} Fermions  & 0.55 & 0.87 & 0.91 & 0.89 & 0.87 &
0.85 & 0.84 & 0.82\\ 
\hspace*{1mm} Gauge Bosons  & 0.23 & 0.69 & 0.91 & 1.00 & 1.04 &
1.06 & 1.06 & 1.07
\\ \botrule
\end{tabular}}
\end{table}


\section{Hawking Radiation in the Bulk}

We now turn to the study of the Hawking radiation emitted by a small,
higher-dimensional black hole in the bulk. According to the assumptions
of the model, only gravitons and scalar fields can propagate
in the space transverse to the brane, therefore, these are the only
modes that should be considered here. Although this type of emission
is unlikely to be observed, the study of the bulk emission is extremely
important since it determines the amount of energy lost in the bulk,
and subsequently the amount of energy left for emission in the form of
brane-localized, observable modes.

Due to the simplicity of the metric, the emission of modes from a
spherically-symmetric, higher-dimensional black hole was studied first
\cite{kmr1}. In addition, the absence of any technical difficulties in
the derivation of the equation of motion for scalar fields propagating
in a higher-dimensional spacetime naturally led to the study of this
particular type of modes first. The emission of bulk scalar modes by a
Schwarzschild-like black hole has been studied both analytically
\cite{kmr1,FS2} and numerically \cite{HK}. We start by reviewing these
results before presenting the definite answer to the question of
bulk-to-brane relative emissivity raised above, and before commenting
further on the emission of scalar and graviton modes in the bulk. 

The gravitational background around a $(4+n)$-dimensional,
spherically-symmetric black hole is described by the line-element
(\ref{metric-n})\,\cite{Myers}, which we give again here for convenience
\begin{equation}
ds^2 = - h(r)\,dt^2 +h(r)^{-1}\,dr^2 + r^2\,d\Omega_{2+n}^2\,.
\label{metr-bulk}
\end{equation}
The equation of motion of a scalar field propagating in the above background
is simply given by $D_M D^M \phi=0$, where $D_M$ is the covariant derivative
in the higher-dimensional spacetime. This equation takes a separable form if
we use the ansatz
\begin{equation}
\phi(t,r,\theta_i,\varphi)=
e^{-i\omega t}\,R_{\omega \ell}(r)\,{\tilde Y}_\ell(\Omega)\, ,
\end{equation}
where ${\tilde Y}_\ell(\Omega)$ is the $(3+n)$-spatial-dimensional
generalization of the usual spherical harmonic functions depending on
the angular coordinates~\cite{Muller}. The radial equation then takes
the form\,\cite{kmr1}
\begin{equation}
\frac{h(r)}{r^{n+2}}\,\frac{d \,}{dr}\,
\biggl[\,h(r)\,r^{n+2}\,\frac{d R}{dr}\,\biggr] +
\biggl[\,\omega^2 - \frac{h(r)}{r^2}\,\ell\,(\ell+n+1)\,\biggr] R =0 \,.
\label{scalareqn}
\end{equation}

As in the case of emission on the brane, the above equation must be solved 
and the absorption coefficient must be determined before we are able to
write down the expression for the greybody factor and the emission rates.
The same analytic method was also used in this case\,\cite{kmr1}:
Eq. (\ref{scalareqn}) was solved in the near-horizon and far-field regime
and the two solutions were matched in an intermediate zone.  In the
near-horizon regime, the same change of variable, $r \rightarrow h(r)$,
brings the radial equation to the form
\begin{equation}
h\,(1-h)\,\frac{d^2R}{dh^2} + (1-h)\,\frac{d R}{dh} +
\biggl[\,\frac{(\omega r_H)^2}{(n+1)^2 h (1-h)} -
\frac{\ell\,(\ell + 1 +n)}{(n+1)^2 (1-h)}\,\biggr] R=0\,,
\label{NH-ell-1}
\end{equation}
while the redefinition $R(h)=h^\alpha (1-h)^\beta F(h)$ reduces it to a
hypergeometric equation with $a=b=\alpha + \beta$ and $c=1+2 \alpha$, where
\begin{equation}
\alpha_{\pm} = \pm \frac{i \omega r_H}{n+1}\,, \qquad
\beta_{\pm} =\frac{1}{2} \pm \frac{1}{n+1}\,
\sqrt{\Bigl(l +\frac{n+1}{2}\Bigr)^2 -(\omega r_H)^2}\,.
\label{al-be-2}
\end{equation}
Then, the general solution of Eq. (\ref{NH-ell-1}) is written as~\cite{AS}
\begin{eqnarray}
&& \hspace*{-0.5cm} R_{NH}(h)=A_- h^{\alpha_{\pm}}\,(1-h)^\beta\,F(a,b,c;h)
\nonumber \\[2mm] && \hspace*{2cm} +\,
A_+\,h^{-\alpha_{\pm}}\,(1-h)^\beta\,F(a-c+1,b-c+1,2-c;h)\,.
\label{NH-gen-2}
\end{eqnarray}
If we expand the above solution in the limit $r \rightarrow r_H$, or
$h \rightarrow 0$, and choose $\alpha=\alpha_-$, we obtain the
approximate form
\begin{equation}
R_{NH} \simeq A_-\,\exp\bigl(-i \omega r_H^{n+2} y\bigr) +
A_+\,\exp\bigl(i \omega r_H^{n+2} y\bigr)\,,
\label{NH-nh}
\end{equation}
in terms of the tortoise coordinate (\ref{tortoise}). Had we chosen the
alternative option $\alpha=\alpha_+$, the same form would have been derived
with the coefficients $A_-$ and $A_+$ interchanged. Since the two options 
are completely equivalent, we choose the former one, and we demand $A_+=0$
in order to have a purely incoming mode near the horizon of the black hole.
The criterion of the convergence of the hypergeometric function
${\bf Re}\,(c-a-b)>0$ demands once again that $\beta=\beta_-$. 

In the far-field zone, the limit $r \gg r_H$ and the redefinition
$R(r)=f(r)/r^{(n+1)/2}$ reduce Eq. (\ref{scalareqn}) to a Bessel
differential equation of the form\,\cite{kmr1}
\begin{equation}
\frac{d^2 f}{dr^2} + \frac{1}{r}\, \frac{d f}{d r} + \Bigl[\,\omega^2 
- \frac{1}{r^2}\,\Bigl(\ell + \frac{n+1}{2}\Bigr)^2\,\Bigl] f=0\,.
\end{equation}
The general solution then for the radial function $R(r)$ is given by
\cite{AS}
\begin{equation}
R_{FF}(r) = \frac{B_+}{r^{(n+1)/2}}\,J_{\ell + (n+1)/2}(\omega r) +
\frac{B_-}{r^{(n+1)/2}}\,Y_{\ell + (n+1)/2}(\omega r)\,,
\label{FFscalar-ell}
\end{equation}
where $J$ and $Y$ are the Bessel functions of the first and second kind,
respectively.

In order to match the two asymptotic solutions, the far-field one is expanded
in the limit $\omega r\ll 1$, giving
\begin{equation}
R_{FF}(r) \simeq \frac{B_+ \,r^\ell}{\Gamma(\ell + \frac{n+3}{2})}
\left(\frac{\omega}{2}\right)^{\ell + (n+1)/2}
- \frac{B_-}{r^{\ell+n+1}} \left(\frac{2}{\omega}\right)^{\ell+(n+1)/2}
\frac{\Gamma(\ell+\frac{n+1}{2})}{\pi}\,,
\label{FFsmall-ell}
\end{equation}
while the near-horizon solution, after being ``shifted" and its argument
changed to $1-h$ \cite{AS}, is expanded in the limit $r \gg r_H$ thus
assuming the form
\begin{equation}
R_{NH}(h) \simeq A_-\,\Gamma(1+2\alpha)\,\biggl[\Bigl(\frac{r}{r_H}
\Bigr)^\ell\,\frac{\Gamma(1-2\beta)}{\Gamma(1+\alpha-\beta)^2}
+ \Bigl(\frac{r_H}{r}\Bigr)^{\ell + n +1}\,
\frac{\Gamma(2\beta-1)}{\Gamma(\alpha+\beta)^2}\biggr]\,.
\label{NHlarger-ell}
\end{equation}
Matching the two solutions (\ref{FFsmall-ell}) and (\ref{NHlarger-ell}), 
we obtain a relation between the two integration constants at infinity
\begin{equation}
\frac{B_+}{B_-} = -\biggl(\frac{2}{\omega r_H}\biggl)^{2\ell+n+1}\,
\frac{\Gamma(\ell +\frac{n+1}{2})^2\,\Bigl(\ell +\frac{n+1}{2}\Bigl)\,
\Gamma(1-2\beta)\,\Gamma(\alpha+\beta)^2}{\pi\,\Gamma(1+\alpha-\beta)^2\,
\Gamma(2\beta-1)}\,. \label{B}
\end{equation}

After having completed the determination of the solution for the radial
function $R(r)$, we turn our attention to the form of the scalar field
at infinity. We need to determine the amplitudes of the incoming and outgoing
modes, thus, we expand Eq. (\ref{FFscalar-ell}) in the limit $r \rightarrow
\infty$, and we find:
\begin{equation}
\label{far-bulk}
R^{(\infty)}=A_{in}^{(\infty)}\,\frac{e^{-i\omega r}}{\sqrt{r^{n+2}}}+
A_{out}^{(\infty)}\,\frac{e^{i\omega r}}{\sqrt{r^{n+2}}}\,,
\end{equation}
with
\begin{eqnarray}
&& A_{in}^{(\infty)} = \frac{B_++ i B_-}{\sqrt{2 \omega \pi}}\,
\,e^{i \pi(\ell + \frac{n}{2}+1)/2}\,, \\[2mm]
&& A_{out}^{(\infty)}=\frac{B_+ - i B_-}{\sqrt{2 \omega \pi}}\,
\,e^{-i \pi(\ell + \frac{n}{2}+1)/2}\,.
\end{eqnarray}
The reflection coefficient ${\cal R}_\ell$ is defined as the ratio of the
outgoing amplitude over the incoming one at infinity. Then, the absorption
coefficient ${\cal A}_\ell$ is given by
\begin{equation}
|{\cal A}_\ell|^2 = 1 - |{\cal R}_\ell|^2 = 
1- \biggl|\frac{B_+ -iB_-}{B_+ + iB_-}
\biggr|^2= \frac{2i\,(B^*-B)}{B B^* + i(B^*-B) +1}\,,
\label{abs-coef}
\end{equation}
where $B \equiv B_+/B_-$ is defined in Eq. (\ref{B}). 

The above analytic result can take a simplified form in the low-energy limit
$\omega r_H \ll 1$, in which case $B B^* \gg i(B^*-B) \gg 1$, and we may write
\begin{equation}
|{\cal A}_\ell|^2= \frac{4 \pi^2}{2^{4 \ell/(n+1)}}\,
\biggl(\frac{\omega r_H}{2}\biggl)^{2\ell+n+2}\,
\frac{\Gamma\Bigl(1+\frac{\ell}{n+1}\Bigr)^2}
{\Gamma\Bigl(\frac{1}{2}+\frac{\ell}{n+1}\Bigr)^2\,
\Gamma\Bigl(\ell +\frac{n+3}{2}\Bigr)^2} + ...\,.
\label{siresult}
\end{equation}
Substituting the above result into Eq. (\ref{greyb}), we arrive
at\,\cite{kmr1}
\begin{equation}
\sigma_{\ell,n} (\omega) = \frac{\pi}{2^{4 \ell/(n+1)}}\,
\biggl(\frac{\omega r_H}{2}\biggl)^{2\ell}\,
\frac{\Gamma\Bigl(1+\frac{\ell}{n+1}\Bigr)^2\,
\Gamma\Bigr(\frac{n+3}{2}\Bigr)^2}
{\Gamma\Bigl(\frac{1}{2}+\frac{\ell}{n+1}\Bigr)^2\,
\Gamma\Bigl(\ell +\frac{n+3}{2}\Bigr)^2}\,N_\ell\,A_H + ...\,,
\label{simple-gb}
\end{equation}
where $N_\ell$ is the multiplicity of states with the same angular momentum
number $\ell$ defined in Eq. (\ref{bulk-mult}). Since $\omega r_H \ll 1$,
the greybody factor decreases as $\ell$ increases, therefore, the main
contribution to $\sum_\ell \sigma_{\ell,n}$ comes from the lowest partial
wave with $\ell=0$.
It is easy to see that the above expression evaluated for $\ell=0$ simply
reduces to $A_H$, thus, revealing the fact that even in the higher-dimensional
case, the greybody factor for scalar fields at the low-energy regime is given
by the area of the horizon \footnote{This has been recently shown 
\cite{Jung} to hold also for massive scalar particles propagating in a
higher-dimensional spacetime.}. This behaviour is similar to the one
obtained in the four-dimensional case; here, however, the area of the
horizon changes as $n$ varies. 

As in the four-dimensional case, the contribution to the greybody factor
from the dominant partial wave comes out to be independent of the number
of extra dimensions. Looking at the dependence of the higher partial waves
on $n$, we obtain a suppression of the greybody factor as the dimensionality
of the bulk increases. However, in order to be absolutely certain about this
behaviour we would have to include next-to-leading-order corrections in the
simplified expression of $\sigma_{\ell,n}(\omega)$ (\ref{simple-gb}), or
simply deal with the full analytic result derived from Eq. (\ref{abs-coef}). 
In either case, however, the derived dependence would only hold in the
low-energy regime and no information could be derived from these expressions
for the dependence of the greybody factor, and thus of the emission rates,
in the high-energy regime.

\begin{figure}[t]
\begin{center}
\begin{tabular}{c}
\epsfig{file=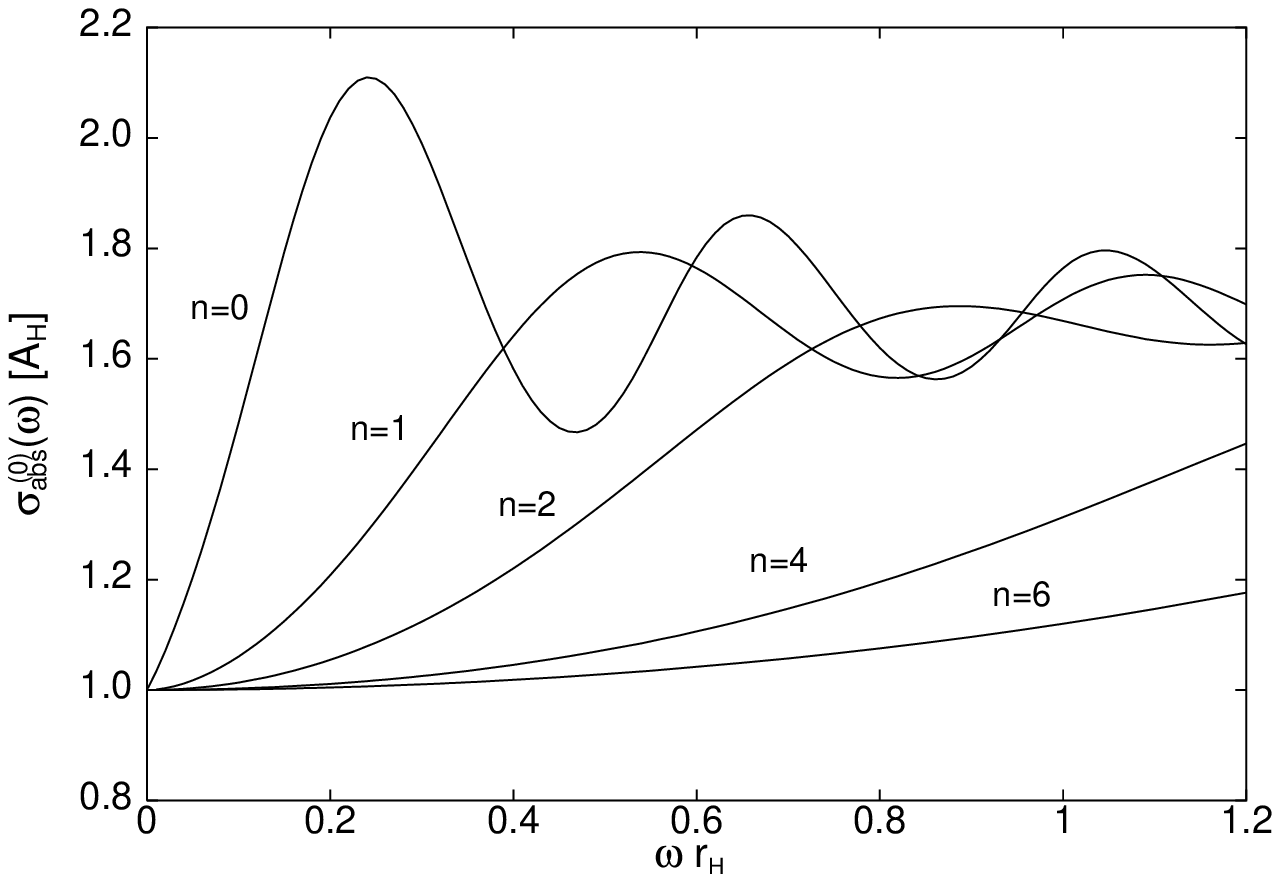,width=6cm} \hspace*{0.1cm}
\epsfig{file=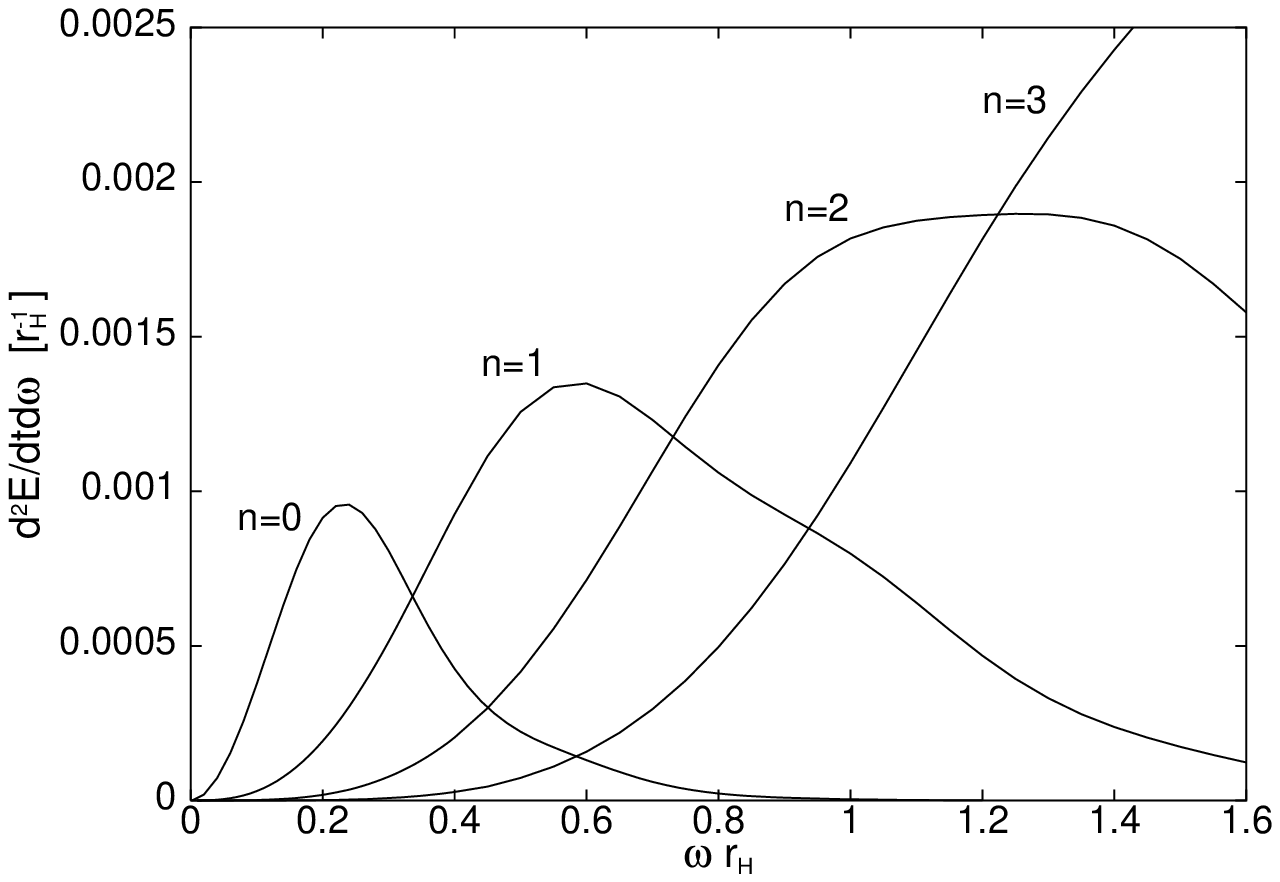,width=6cm}\end{tabular}
\end{center}
\caption{{\bf (a)} Greybody factors and {\bf (b)} energy emission rates for
scalar emission in the bulk from a $(4+n)$-dimensional black hole.}
\end{figure}

For this reason, we turn to the numerical integration of the radial equation
(\ref{scalareqn}) that can provide exact results for the greybody factor
and energy emission rate in the bulk valid at all energy regimes. The
behaviour of both these quantities are shown in Figures 6(a,b) \cite{HK}.
As it was predicted by the analytic expression, the greybody factors,
normalized to the area of the horizon, tend to unity for all values of $n$,
at the low-energy regime. The naive prediction, also coming from the
analytic expression, for the suppression of the greybody factor with $n$,
at the same energy regime, is verified. Similarly to the emission
on the brane, as the energy parameter $\omega r_H$ increases, the greybody
factor oscillates around an asymptotic high-energy limiting value. In
analogy with the four-dimensional case,  we expect the greybody factor to
adopt the $(4+n)$-dimensional geometrical optics limit at the high-energy
regime. For large values of the energy of the scattered particle, the
greybody factor becomes equal to the area of an absorptive body of radius
$r_c$ which is projected on a plane parallel to the one of the orbit of
the moving particle \cite{MTW}. We may compute this area by setting one
of the azimuthal angles $\theta_i$ equal to $\pi/2$ and integrating over 
the remaining angular coordinates. Then, we find \cite{HK}
\begin{equation}
A_p = \frac{2 \pi}{(n+2)}\,\frac{\pi^{n/2}}{\Gamma[(n+2)/2]}\,
r_c^{n+2}\,.
\end{equation}
The area of the absorptive body is clearly $n$-dependent, and reduces, as
expected, to the usual four-dimensional result $A_p = \pi r_c^2$, for $n=0$.
In the above, $r_c$ is the value of the effective radius defined in
Eq. (\ref{effective}) \cite{emparan}. The greybody factor then is given
by the above projected area, or more explicitly,
\begin{equation}
\sigma_{g}= 
\frac{1}{\sqrt{\pi}\,(n+2)}\,\frac{\Gamma[(n+3)/2]}{\Gamma[(n+2)/2]}\,
\biggl(\frac{n+3}{2}\biggr)^{(n+2)/(n+1)}\,
\biggl(\frac{n+3}{n+1}\biggr)^{(n+2)/2}\,A_H\,.
\label{high}
\end{equation}
In Ref. \refcite{HK}, it was pointed out that the high-energy asymptotic
value of the greybody factor, determined through numerical integration,
although very close to the ones following from the above analytic expression,
did not exactly coincide. However, further study of the asymptotic behaviour
revealed that the numerical results do indeed agree with the analytic
prediction. The apparent disagreement was due to the fact that the bulk
greybody factor adopts its asymptotic value at much higher energies than
the brane greybody factor -- extending our numerical calculation to
higher energies led to the complete agreement with Eq. (\ref{high}).

According to Figure 6(b), the emission rate of scalar fields in the bulk is
enhanced as the number of extra dimensions increases. As with the emission
on the brane, this is caused by the increase in the temperature of the black
hole, which eventually overcomes the decrease in the value of the greybody
factor and causes the enhancement of the emission rate with $n$ at high
energies. We can now provide an accurate answer to the question of how much
energy is emitted into the bulk compared to the one on the brane. By using
Eq. (\ref{alter-bulk}), the relative bulk-to-brane energy emission rate
may be written as
\begin{equation}
\frac{d E^{(B)}/dt}{d E^{(b)}/dt} = \frac{\sum_{\ell} N_\ell^{(B)}\,
|{\cal A}_\ell^{(B)}|^2}{\sum_{\ell} N_\ell^{(b)}\,
|{\cal A}_\ell^{(b)}|^2}\,,
\label{ratio}
\end{equation}
where the superscripts $B$ and $b$ denote emission in the bulk and brane,
respectively. In the above, we have taken into account that bulk and brane
modes `feel' the same temperature. The multiplicity of states in the bulk
rapidly increases with $n$, compared to the one on the brane that remains
constant. However, it turns out that the enhancement of the absorption
probability for brane emission, as $n$ increases, is considerably greater
than the one for bulk emission. The exact numerical results derived in
Ref. \refcite{HK} allow us to compute, once again, the total emissivity
of the black hole in the bulk and compare it with the one on the brane.
The total bulk-to-brane relative emissivities, as a function of $n$, 
are given in Table 6\,\cite{HK}.

\begin{table}[ht]
\tbl{Bulk-to-brane energy emissivities for different values of $n$}
{\begin{tabular}{@{}ccccccccc@{}} \toprule
$n$ & 0 & 1 & 2 & 3 & 4 & 5 & 6 & 7 \\ \colrule
\hspace*{1mm} {\rm Bulk/Brane} & 1.0 & 0.40 & 0.24 & 0.22 & 0.24
& 0.33 & 0.52 & 0.93 \\ \botrule
\end{tabular}}
\end{table}

In Ref. \refcite{emparan}, it was shown that the whole tower of KK excitations
of a given particle carries approximately the same amount of energy as the
massless zero-mode particle emitted on the brane. Combining this result with
the fact that many more types of particles live on the brane than in the
bulk, it was concluded that most of the energy of the black hole goes into
brane modes. The results obtained in Ref. \refcite{emparan} were only
approximate since the dependence of the greybody factor on the energy of
the emitted particle was ignored and the (low-energy only valid) geometric
expression for the area of the horizon was used instead \footnote{In Ref.
\refcite{cavaglia}, it was correctly argued that the dependence of the
greybody factor on $n$ must be taken into account, however, the dependence
on energy $\omega$ was again ignored.}. The numerical results reviewed
above have
used the exact dependence of the greybody factor on both the energy and
number of extra dimensions in order to provide the most accurate estimate
for the bulk-to-brane energy emissivity.
From the entries of the above Table, it becomes clear that the emission
of bulk modes for intermediate values of the number of extra dimensions,
i.e. $n=2,3,4$ and 5, is particularly suppressed compared to the one
of brane modes.  For lower or higher values of $n$, the emission of bulk
modes becomes important but the bulk-to-brane ratio never exceeds
unity. In conclusion, most of the energy of the higher-dimensional black
hole, in the `scalar' channel, is indeed emitted directly on the brane,
nevertheless, depending on the number of extra dimensions, a substantial
amount of the total energy may be lost in the bulk. 

However, a definite conclusion regarding the {\it total} amount of energy
which is lost in the bulk cannot be drawn before the emission of gravitational
radiation is also studied. A master equation describing scalar, vector and
tensor gravitational perturbations, in a higher-dimensional
spherically-symmetric spacetime, was recently derived \cite{KI} and was
shown to take the form of a 2nd order differential wave equation. 
At the moment, the interest has been focused on the calculation
\cite{Cardoso2,Konoplya,Cardoso3,Berti} of the corresponding quasinormal
modes \cite{KS}, resonances at complex frequencies that dominate at late
times after the perturbation of the black hole background (see also
\cite{Horowitz,Konoplya2,Molina,Cardoso4,Siopsis} for the study of quasinormal
modes of scalar fields in the higher-dimensional spacetime). No study
addressing the question of the relative bulk-to-brane emission rate for
gravitons has yet been done.

The emission of Hawking radiation in the bulk during the spin-down phase
has again not been studied as much as the one during the Schwarzschild phase. 
The equation of motion of a scalar field in a five-dimensional Kerr-like
background was shown \cite{FS3} to take a separable form (just like the
Hamilton-Jacobi equations for particles and light propagating in the same
background \cite{FS4}) and analytic formulae for the energy and angular
momentum fluxes, valid at the low-energy regime, were written down; no
quantitative results, however, for the radiation spectra were produced
and no generalization of this analysis for arbitrary number of extra
dimensions has been carried out. The study of the stability of 
higher-dimensional, rotating black holes has also attracted some attention:
in Ref. \refcite{EM} it was argued that `ultra-spinning' ($4+n$)-dimensional
black holes, i.e. with an arbitrarily large angular momentum, are unstable
-- an earlier work had suggested that black holes with an angular momentum 
beyond a critical value will `decay' to a rotating black ring \cite{ER}. On
the other hand, perturbations on a massless scalar field propagating in a
Kerr-like, higher-dimensional black hole was shown \cite{IUM} to be free
of unstable modes, no matter how large the angular parameter $a$ is.
Finally, the stability of a five-dimensional rotating black hole projected
onto the brane was checked \cite{BKP}, by studying scalar, electromagnetic
and gravitational perturbations, and no unstable modes were found in the
spectra. It is worth mentioning at this point that the Schwarzschild-like
higher-dimensional black holes were shown \cite{IK2} to be stable under
vector and tensor gravitational perturbations, with a potential instability
arising only from the scalar gravitational sector, for $n>2$.

\section{Conclusions}

The proposal of the existence of extra dimensions in the Universe has opened
many different pathways that, in principle, lead to important modifications
in the four-dimensional cosmology, particle physics phenomenology and black
hole physics. The recent revival of the idea of the existence of extra
space-like dimensions, that can have an almost macroscopic size or be 
even non-compact, introduced a novel feature in the theory: the scale at
which the gravity becomes strong may be much lower than the traditional
four-dimensional Planck mass $M_P$. This has led to the exciting prospect
that high-energy collisions between elementary particles, that will take
place at next-generation ground-based accelerators or have been already
taken place at the atmosphere of the Earth, can probe the energy regime of
quantum gravity. Some of the most striking consequences would be the
possible creation of mini black holes, or even strings and D-branes, as
the products of a high-energy collision of particles with center-of-mass
energies at just a few times the new scale of gravity $M_*$. 

In Section 2, we have reviewed the existing results in the literature
regarding the possibility of the creation of mini black holes during
high-energy collisions. Well-known, four-dimensional analyses 
have been generalized to cover the case where the colliding particles
propagate in a higher-dimensional spacetime. Some of the new studies have
shown that, in the case of head-on collisions, the mass of the produced
black hole gets suppressed, as the number of extra dimensions increases,
while some recent complimentary results indicate that the emission of
gravitational waves during the collision is actually suppressed when
$n$ increases: this obviously brings the two types of results in an
apparent disagreement unless we accept the possibility that a significant
amount of the energy, lost during the head-on collision, is emitted
in a form different from that of gravitational radiation. On the other
hand, for collisions with a non-vanishing impact parameter, which are the
most likely to take place, the black hole production cross-section is in
fact enhanced with the number of extra dimensions. Putting the above into
the framework of a realistic collision between composite particles have 
led to large estimates, by new physics standards, for the corresponding
black hole production cross section, either at the LHC or at the
atmosphere of the Earth.

The effect of extra dimensions is not restricted to the production of mini
black holes; the properties of the produced, higher-dimensional black holes
are also modified. In Section 3, we have discussed some of those properties,
namely, the horizon radius, temperature and lifetime, all of which are
crucial parameters for the successful production and detection of these
elusive, up to now, objects. As we saw, the horizon radius of a
($4+n$)-dimensional black hole is many orders of magnitude larger than the
one of a four-dimensional black hole with the same mass, which simply means
that a mass $M$ needs to be compacted less in a higher-dimensional spacetime
to create a black hole. The temperature of these small black holes, in turn,
comes out to be lower than in four-dimensions, which means that the emission
rate of Hawking radiation is smaller and their lifetime longer. What is
most favourable for the possibility of detecting these objects is the
fact that, for a black hole with a mass $M_{BH}=5$ TeV or so, the Hawking
radiation spectrum reaches its peak at energies close to the temperature of
the black hole which is of the order of 100-600 GeV (for $n=1,...,7$); 
this is exactly
the energy regime that present and next-generation collider experiments can
probe. Such a small black hole will be extremely short-lived, i.e.
$\tau \simeq 10^{-26}$ sec, nevertheless, the proximity of the evaporating
black holes to our detectors increases significantly the possibility
of their detection.

These mini black holes, upon implementation of quantum effects, emit Hawking
radiation into the higher-dimensional spacetime in the form of both bulk and
brane modes. The generalization of the four-dimensional expressions for the
emission rates in the case of  a higher-dimensional spacetime is
straightforward, nevertheless, the exact expression of the greybody factors
for different types of fields propagating in a higher-dimensional background
was, until recently, unknown. As we explained in Section 4, the greybody
factors encode valuable information for the background around the emitting
black hole and depend on the energy of the emitted particle, its spin and
the dimensionality of spacetime. This means that the presence of the greybody
factor in the radiation spectrum will cause the modification of the
low-energy emission rate from the high-energy one, and
will lead to different emissivities for particles with different spin.
Moreover, both the number and the type of particles emitted will depend
on the number of extra dimensions that exist in nature, a feature
that may possibly lead to the determination of the dimensionality of
spacetime upon detection of Hawking radiation.

Having concluded that the implementation of the greybody factor in the
radiation spectrum is imperative in order to derive accurate estimates
for the emission spectrum of the black hole, we moved on, in Section 5,
to derive a master equation for the propagation of fields with different
spin on the black-hole background induced on the four-dimensional brane. By
solving this master equation, one can compute the transmission cross-section,
in other words the greybody factor, for brane-localized modes emitted
by the black hole. This type of emission during the spin-down phase of a
black hole has been only partially studied: radiation spectra for fields
with different spin $s$ have been computed only for the case of $n=1$ 
in the limit of low energy and low angular momentum. On the contrary, the
spherically-symmetric Schwarzschild phase has been thoroughly investigated.
Both analytical and numerical methods were used, with the former one
leading to analytical, but low-energy-only-valid, expressions, and the
latter one yielding exact numerical results valid at all energy regimes.
We were thus able to compare the radiation spectra for different types of
fields and different number of extra dimensions. For all types of particles,
the total emissivities are greatly enhanced with the number of extra
dimensions, with the enhancement reaching orders of magnitude for large
values of $n$. As the increase in the emission rate depends strongly on
the spin, the relative emissivities for particles with different spin
are also strongly $n$-dependent: while scalar fields remain the preferred
type of particle emitted by the black hole for low and intermediate values
of $n$, they are outnumbered by the gauge bosons for large values of $n$,
with the fermions being the least effective channel during the emission.

The emission of brane-localized modes is undoubtly the most
phenomenologically interesting effect since it involves Standard Model
particles that can be easily detected during experiments. On the other hand,
the emission of bulk modes will be only perceived as a missing energy
signal by the observer on the brane. Nevertheless, the amount of energy
lost in the bulk is crucially important as it determines the remaining
available energy for emission on the brane. The details of the emission
of bulk scalar modes during the spin-down phase have been studied in an
analytic but qualitative way, and no quantitative results are available.
For the Schwarzschild phase, as we saw in section 6, the study has been
completed. The
greybody factor and emission rates have been calculated and the latter
were shown to increase again with the dimensionality of spacetime. 
The extremely important question of the bulk-to-brane energy emissivity
has been answered only for the scalar channel and for the Schwarzschild
phase: the amount of energy spent by the black hole for the emission of
bulk scalar modes is always smaller than the one for brane modes; 
nevertheless, the amount of energy lost in the bulk must always be taken
into account especially for large values of $n$ when the energies spent 
in the brane and bulk channel become comparable. No results for the
bulk-to-brane emissivity for gravitons have been yet derived.

Although the possibility of the production and evaporation of mini black
holes at the LHC is an exciting prospect, this will be possible only in the
case where the fundamental scale of gravity $M_*$ is indeed very close to
1 TeV. Nevertheless, there is absolutely no guarantee for that, and the
only argument in favour of this particular value is the possible resolution
of the hierarchy problem. If $M_*$ is larger than 1 TeV, even by one order
of magnitude, the probability of the production of black holes at the LHC
vanishes (although we might still witness these type of effects in cosmic ray
particles). Nevertheless, all the analytical and numerical results presented
in this review have $M_*$ as an independent parameter, and are therefore
valid for all values of $M_*$. If this scale is pushed upwards by various
constraints, the derived results will still be applicable for the production
and evaporation of black holes at the new energy regime $\sqrt{s} \geq M_*$. 

We would also like to stress that the results for the radiation spectra
reviewed here refer to individual degrees of freedom and not to elementary
particles, like electrons or quarks, which contain more than one polarization.
For the number of elementary particles produced, and the energy they carry,
one has to use a Black Hole Event Generator \cite{dl,HRW}. This simulates
both the production and decay of small black holes at hadronic colliders
and provides estimates for the number and spectra of the different types
of elementary particles produced.

In this review, we concentrated on theories postulating the existence of
Large Extra Dimensions, and we studied the properties of small
black holes that live in a spacetime with $D-1$ flat spacelike dimensions. 
Our analysis is definitely not valid in highly curved spacetimes, like
the ones in five-dimensional warped models~\cite{RS}. In that case,
attempts to  construct a gravitational background that would reduce to a
black-hole line-element on the brane while being well defined away from
it have failed; numerical methods have instead found five-dimensional
localized black holes that do not necessarily have a black hole line-element
projection onto the brane (see Refs. \refcite{CHR}-\refcite{Kudoh} for some
relevant works). Nevertheless, in the case of very small bulk cosmological
constant, the `warping' of the five-dimensional spacetime, parametrised
by the inverse AdS radius, $\lambda^{-1}$, is small and the extra spacetime
can be considered as an almost flat one. Alternatively, if the horizon
radius $r_H$ is much smaller than the AdS radius, no matter what the value 
of $\lambda$ is, then the black hole
cannot perceive the warping of the fifth dimension. In either case, we may
model black holes with $r_H \ll \lambda$ arising in a warped spacetime
as black holes living in a five-dimensional, flat spacetime. Under this
assumption, all the results presented in this review hold equally well
also for the `warped' black holes.

As a concluding remark, we would like to stress once again that the 
detection of signatures of possible black hole production events during 
high-energy collisions would be a revolutionary development both for
particle physics phenomenology and gravitational physics. Any observational
signal of this type would automatically prove the existence of extra
dimensions and of a fundamental theory of all forces, with a low energy
scale, valid in a higher-dimensional spacetime. The detection of Hawking
radiation emitted by these small, higher-dimensional black holes is the
most direct evidence for the production and evaporation of those objects.
The radiation spectrum of a decaying black hole can also reveal the exact
dimensionality of spacetime as both
the amount and  type of the emitted radiation strongly depend on it. What
is also exciting is that a black hole can emit all types of particles that
exist in nature, independently of their spin, charge, quantum numbers or
interaction properties, as long as their rest mass is smaller than the
black hole temperature; therefore, long-sought but yet undiscovered
particles, like the Higgs fields or supersymmetric particles, might
indeed make their appearance in the decay spectrum of a black hole.
The launch of the LHC, or of any other future experiment able to probe 
even higher energy regimes, deserves to be awaited with great expectations
indeed.

\section*{Acknowledgments}

I would like to thank John March-Russell and Chris. M. Harris for
constructive and enjoyable collaborations. This work was funded by the
U.K. Particle Physics and Astronomy Research Council (Grant Number
PPA/A/S/2002/00350).

\end{document}